\documentclass{amsart}
\usepackage{orcidlink}
\usepackage{url}
\usepackage{graphicx} 
\usepackage{amsthm}
\theoremstyle{definition}
\newtheorem{definition}{Definition}
\newtheorem{example}{Example}
\usepackage{comment}
\usepackage[margin=1 in]{geometry}
\usepackage{float}
\title{Ballot Exhaustion in Multiwinner Single Transferable Vote Elections}
\author{David McCune \and E.E. Naber}

\begin{document}

\begin{abstract}
We study ballot exhaustion in multiwinner single transferable vote (STV) elections using a dataset of 1,070 Scottish local government elections comprising over 5.4 million ballots. While ballot exhaustion has been studied extensively in single-winner elections, comparatively little work examines exhaustion in the multiwinner setting. We introduce formal definitions of several types of exhaustion in STV elections, distinguishing between exhausted ballots, non-first-choice exhausted ballots, unrepresented exhausted ballots, and weight exhaustion. These definitions clarify important conceptual differences between ballots that cease to transfer and ballots that fail to contribute meaningfully to representation.

Our empirical analysis shows that 27.9\% of ballots are exhausted by the final round of counting, although the corresponding weight exhaustion rate is only 7.1\%, indicating that many exhausted ballots have already contributed to the election of a candidate. Moreover, most exhausted ballots correspond to voters who achieve some form of representation, either because their first-ranked candidate wins or because a candidate ranked among their top choices is elected. These results suggest that raw exhaustion rates alone substantially overstate the extent to which voters lose their influence or fail to obtain representation under STV. We also investigate whether exhaustion can affect electoral outcomes by extending partial ballots under several completion models. Under extreme assumptions, exhaustion can potentially alter a substantial number of outcomes, but under a proportional ballot-completion model only 3.5\% of seats change. Finally, we show that a substantial number of winners fail to reach quota, even after the elimination of all losing candidates. These results help clarify the practical and normative significance of ballot exhaustion in real-world STV elections.  
\end{abstract}

\maketitle

\section{Introduction}
Ranked-choice voting methods such as the single transferable vote (STV) are often justified on the grounds that they produce representative outcomes by allowing voters to express a ranking of the candidates. However, in practice many voters do not rank all candidates and, as a result, some ballots cease to influence the outcome as the counting unfolds. This phenomenon, commonly referred to as \emph{ballot exhaustion}, raises important questions about representation and a voter's ``loss of influence.'' In this article we examine ballot exhaustion in multiwinner STV elections from Scotland, expanding on the work of Endersby and Towle \cite{ET14}, who study STV elections in Ireland. 

Ballot exhaustion has been primarily studied in the single-winner case, with empirical work dating to Burnett and Kogan \cite{BK15}, who study four American ranked choice elections. They observe that due to the prevalence of partial ballots (ballots which do not provide a complete ranking of all candidates), many voters have no influence over the choice of the winner in the final round of vote tabulation, and sometimes a candidate can win without securing an electoral majority. Graham-Squire and McCune \cite{GSM25} build on this analysis by examining 185 American ranked choice elections, finding similar results. McCarty \cite{M24} analyzes how ballot exhaustion affects voters of different ethnic and racial minorities, finding ``strong evidence that electorates with heavy concentrations of ethnic and racial minorities have substantially higher rates of ballot exhaustion.'' Other studies \cite{DMM, KGF, TUK} analyze exhaustion obliquely by investigating the extent to which the presence of partial ballots can affect the election winner.

The phenomenon of ballot exhaustion is closely connected to the issue of ballot errors, as both reflect limitations in how voters express ranked preferences and can lead to ballots that are only partially effective—or not counted at all—in determining the outcome. Cormack \cite{C23} and Pettigrew \cite{P25} both study the frequency with which voters make errors when filling out ranked ballots, often causing the ballots not to be counted. Both studies find a high rate of ballot errors, as compared to plurality ballots which ask a voter to vote for only a single candidate. 

Previous work on ballot exhaustion in multiwinner elections is limited. Endersby and Towle \cite{ET14} study ballot exhaustion in multiwinner Irish STV elections but their study, while very substantive, is limited by a lack of access to actual ballot data. Furthermore, Endersby and Towle do not use the language of exhaustion, but instead focus on ballot non-transferability and ``wasted votes.'' Deshpande et al. \cite{DGJ26} analyze exhaustion in four multiwinner STV elections in Portland, Oregon, but these four elections are not the primary focus of their work; instead, they mainly analyze single-winner ranked-choice elections. Prior theoretical work acknowledges that exhaustion can arise under STV (see \cite{T95}, for example) but there has been little systematic treatment or empirical investigation of the phenomenon.

Our work makes two main contributions. First, we provide formal definitions of several types of ballot exhaustion in the multiwinner STV setting, clarifying distinctions that are sometimes conflated in prior work. Second, we conduct a large-scale empirical analysis of ballot exhaustion using a dataset of 1,070 Scottish local government elections, comprising over 5.4 million ballots. For these elections we have access to actual ballot data, allowing us to greatly expand the analysis of Endersby and Towle \cite{ET14}.

Our results show that while a substantial fraction of ballots become exhausted, the proportion of total vote weight lost is much smaller, highlighting an important distinction between ballot-level and weight-based measures. Moreover, although ballot exhaustion can in principle affect electoral outcomes under extreme assumptions, we find that under more realistic models of voter behavior its impact is limited, typically affecting only a small fraction of seats. STV is often defended on the grounds that ranked ballots preserve voter voice through transfers, but this promise depends critically on voters ranking enough candidates for those transfers to occur. Understanding how often ballots cease to influence outcomes, and in what sense, is therefore central to evaluating whether STV achieves its representative goals in practice.

A major theme of this article is that the democratic significance of ballot exhaustion depends on how exhaustion is defined. While raw exhaustion rates in multiwinner STV elections can appear quite large, many exhausted ballots have already contributed to the election of a candidate before becoming inactive. As a result, different notions of exhaustion can lead to substantially different interpretations of how much voter influence is ultimately lost during the counting process. While ballot exhaustion and its consequences are legitimate concerns for STV, our results show that many exhausted ballots nevertheless contribute meaningfully to representation, and that the practical effect of exhaustion on electoral outcomes appears limited in most cases.

The article is structured as follows. Section \ref{section:STV_description} defines the version of STV implemented in Scotland, and Section \ref{section:elections_description} describes the Scottish dataset. Section \ref{section:exhaustion} provides our definitions of types of exhaustion, as well as a discussion of why previous researchers find exhaustion concerning. Section \ref{section:results} gives all of our results, which are analyzed in our Discussion section (Section \ref{section:discussion}). Section \ref{section:conclusion} concludes, and the Appendix contains most of our figures.

\section{Single Transferable Vote}\label{section:STV_description}

STV is a family of voting methods,  where $n$ candidates compete for $S \le n$ legislative seats (i.e., $S$ is the number of winners).  We provide a broad outline of the STV rules in Scotland since Scottish elections are our focus, and for the rest of the article when we reference ``STV'' we mean the STV rules as implemented in Scotland. Many other forms of STV have been proposed, such as Meek \cite{M} or Warren \cite{W} STV. See \url{https://opavote.com/methods/single-transferable-vote} for a good overview of the different STV procedures which have been proposed and implemented. In the special case where $S=1$, we refer to STV as ``instant runoff voting'' (IRV).

In an STV election, each voter casts a ballot expressing a preference ranking of the candidates; when describing ballots, we use the symbol $\succ$ to denote that a voter prefers one candidate to another. For example, if $n=5$ then the ballot $A\succ B \succ C \succ D \succ E$ indicates that this voter ranks candidate $A$ first, $B$ second, etc. In most real-world STV elections voters are not required to provide a complete ranking, so a voter could cast the ballot $A \succ B$, for example. We define the \emph{ballot length} of a ballot to be the number of candidates ranked on that ballot. For example, the ballot $A \succ B$ has length two, while a complete ranking of all $n$ candidates has length $n$. 

Once all ballots are cast they are aggregated into a \emph{preference profile}, which shows how many voters cast each type of ballot. Table \ref{table:AK_election} provides the preference profile for the 2022 Special House election in Alaska, an IRV election where the three candidates Nick Begich (B), Sarah Palin (Pa), and Mary Peltola (Pe) competed for one seat. The number 27070 denotes that this number of voters cast the ballot $B\succ Pa\succ Pe$, for example.

\begin{table}
\centering
\begin{tabular}{ll|ccccccccc}
\multicolumn{2}{c|}{Num. Voters} 
& 27070 & 15478 & 11262 
& 34078 & 3659 & 21237 
& 47407 & 4647 & 23733 \\
\hline
\multicolumn{2}{c|}{1st choice}  
& $B$ & $B$ & $B$ 
& $Pa$ & $Pa$ & $Pa$ 
& $Pe$ & $Pe$ & $Pe$ \\
\multicolumn{2}{c|}{2nd choice}  
& $Pa$ & $Pe$ &  
& $B$ & $Pe$ &  
& $B$ & $Pa$ &  \\
\multicolumn{2}{c|}{3rd choice}  
& $Pe$ & $Pa$ &  
& $Pe$ & $B$ &  
& $Pa$ & $B$ &  \\

\end{tabular}

\caption{The preference profile for the 2022 Alaska Special House election. Table taken from \cite{GSM23}.}
\label{table:AK_election}
\end{table}

\begin{table}
\begin{tabular}{ccc}
\begin{tabular}{l|cc}
Candidate & Round 1 &Round 2\\
\hline
Begich & 53810 & \\
Palin &  58974 & 86044\\
Peltola & 75787&\textbf{91265}\\
\end{tabular}
&&
\begin{tabular}{l|cc}
Candidate & Round 1 &Round 2\\
\hline
Begich & 53810 &\textit{61869.19} \\
Palin &  58974 & 59763.99\\
Peltola & \textbf{75787}&\\
\end{tabular}
\end{tabular}
\caption{(Left) The votes-by-round table for the preference profile in Table \ref{table:AK_election} and $S=1$. (Right) The votes-by-round table for the preference profile in Table \ref{table:AK_election} and $S=2$.}
\label{table:AK_votes_tables}
\end{table}

The STV method takes as input a pair $(P,S)$, where $P$ is a preference profile, and outputs a set of winners $W(P,S)$ of size $S$ (it is possible for the STV process to output multiple winner sets due to ties).  To compute the winner set, STV proceeds in rounds where in a given round either a candidate is eliminated from the election because they have the fewest first-place votes, or a candidate is elected to fill one of the $S$ seats because they received enough first-place votes. To earn a seat, a candidate's first-place vote total must reach the election's \emph{integer Droop quota}, which is defined by \[\text{quota } = \left\lfloor \frac{\text{Number of Voters}}{S+1}\right\rfloor +1.\]

In a given round, if no candidate's first-place vote total is at least the quota then the candidate with the fewest first-place votes is eliminated, and this candidate's votes are transferred to the next candidate on their ballots who has not been elected or eliminated. If a candidate's first-place vote total is greater than or equal to the quota, that candidate is elected and the votes they receive above quota (the candidate's \emph{surplus votes}) are transferred proportionally to the next non-eliminated and non-elected candidate which appears on the ballots being transferred. That is, if a candidate $A$ achieves quota then the first-place votes that $A$ has earned in order to reach quota are ``locked in,'' and only $A$'s surplus votes are transferred proportionally to other candidates, typically resulting in fractional votes.\footnote{Other systems use different means of transferring. The Irish system uses whole-ballot transfers, as discussed in Section \ref{section:exhaustion_IRV_subsection}. Some systems, such as Hare STV, introduce randomness by transferring some ballots at full value, instead of all ballots at fractional value.} If more than one candidate achieves quota in a given round, the candidate with the largest vote total has their surplus transferred in that round. The method continues in this fashion until $S$ candidates are elected, or until some number $S'<S$ of candidates have been elected by surpassing quota and there are only $S-S'$ candidates remaining who have not been elected or eliminated.  The precise rules used in Scottish elections can be found at \url{https://www.legislation.gov.uk/sdsi/2007/0110714245}.

In the next example, we illustrate the STV algorithm for the preference profile in Table \ref{table:AK_election} for $S\in\{1,2\}$.
\begin{example}\label{example:AK_election}

The election contains 188,571 voters, and thus when $S=1$ the quota is a simple majority of 94,286 votes, and when $S=2$ the quota is 62,858. Regardless of the value of $S$, the STV algorithm begins by counting each candidate's first-place votes. 

The first place vote-totals for Begich, Palin, and Peltola are 53810, 58974, and 75787, respectively. When $S=1$ no candidate reaches quota after the initial round of counting, and Begich is eliminated. As a result, ballots of the form $B\succ Pa \succ Pe$ become $Pa \succ Pe$ and 27070 votes are transferred to Palin; Begich is similarly deleted from ballots of the form $B\succ Pe \succ Pa$ and 15478 ballots are transferred to Peltola. After the transfer Peltola has secured a majority of the remaining votes and is declared the IRV winner.

Because preference profiles are often too large to display in practice, election offices publish election results in the form of a \emph{votes-by-round} table, which shows the number of first-place votes for each candidate in each round of counting. The left of Table \ref{table:AK_votes_tables} shows this table for $S=1$.

When $S=2$, Peltola achieves quota in the first round and a ballot with her ranked first is  given weight $(75787-62858)/(47407+4647+23733)=0.17$, and Peltola is removed from all ballots. Thus, a ballot of the form $Pe\succ B \succ Pa$ is transformed to $B\succ Pa$ and $47407\cdot 0.17=8059.19$ votes are transferred to Begich. Similarly, $4647\cdot 0.17=789.99$ votes are transferred to Palin, and $23733\cdot 0.17=4034.61$ become empty and are transferred to no one. The right of Table \ref{table:AK_votes_tables} shows the votes-by-round table for the $S=2$ election. Note the effect of this ballot re-weighting is that Peltola retains the 62858 votes needed to achieve quota, and a maximum of 12929 votes (her surplus) can be transferred to other candidates.
\end{example}
This example motivates the idea of ballot exhaustion, the key notion for our paper. In both elections, some ballots become empty in the final round because they no longer rank a remaining candidate. However, in the $S=2$ election this may be less concerning because voters who ranked Peltola first already secured representation through the election of their top-ranked candidate.

 \section{Scottish Local Government Elections}\label{section:elections_description}

Scottish local government elections have used STV since 2007, replacing the previously used first-past-the-post system. Our dataset consists of 1,070 ward-level elections from the 2007, 2012, 2017, and 2022 election cycles, comprising a total of 3,718 seats and 5,483,680 ballots. Scottish local elections provide an especially informative setting for studying ballot exhaustion because they combine large-scale use of STV, voluntary partial ranking, and publicly available preference-profile data. Unlike many STV jurisdictions, the Scottish system uses fractional transfers and produces publicly available ballot data that allow individual ballots to be tracked round-by-round throughout the counting.

Scotland is partitioned into 32 council areas, governed by elected councils. These areas are further divided into wards, which elect a fixed number of councilors to represent them. The number of councilors per ward is determined primarily by population, although other factors also play a role. Elections are held every five years, with all seats in a ward filled using STV. Preference profiles from the 2012, 2017, and 2022 election cycles were collected for the work in \cite{MGS} and are publicly available at \url{https://github.com/mggg/scot-elex}. The repository also contains some single-winner IRV elections, which we exclude due to our focus on multiwinner elections. Preference profiles from the 2007 elections are difficult to obtain; the only profiles available from this election cycle are from the Glasgow City council area and can be accessed at preflib.org \cite{MW}.


Table \ref{table:Scottish_data_summary} shows the number of elections in the dataset for each pair $(n,S)$. The table shows that most elections have three or four seats, and typically $n \in \{5,6,7,8,9\}$. Since these are local elections, the electorates are relatively small: across all elections the minimum number of voters is 661, the maximum is 14,207, and the median is 4,863.

The preference profiles from these Scottish elections provide a very rich source of ranked ballot data. These profiles have been used in a range of studies, including estimating the frequency of monotonicity paradoxes \cite{MGS}, evaluating how well STV produces proportional outcomes \cite{BBMP24, BDDW24, M23}, and comparing STV outcomes to those of other voting methods \cite{MMLS24}. Our work in this article builds on this literature by focusing specifically on ballot exhaustion.

In addition to ballot-level analyses, the use of STV in Scotland has also been studied from a political and institutional perspective. For example, Clark analyzes how voters and parties have adapted to STV in local elections \cite{C13, C21}, while Curtice \cite{C22} examines party performance and nomination strategies.

\begin{table}
\begin{tabular}{c|cccccccccccc|c}

$S$ \textbackslash \space $ n$ & 3 & 4 & 5 & 6 & 7 & 8 & 9 & 10 & 11 & 12 & 13 & 14 & Total \\
\hline
2 & 2 & 1 & 0 & 0 & 0 & 1 & 0 & 1 & 0 & 1 & 0 & 0 & 6 \\
3 & 0 & 35 & 76 & 146 & 156 & 90 & 36 & 10 & 3 & 0 & 1 & 0 & 553 \\
4 & 0 & 0 & 34 & 58 & 125 & 115 & 82 & 59 & 22 & 7 & 5 & 1 & 508 \\
5 & 0 & 0 & 0 & 0 & 0 & 1 & 0 & 0 & 1 & 1 & 0 & 0 & 3 \\
\hline
Total & 2 & 36 & 110 & 204 & 281 & 207 & 118 & 70 & 26 & 9 & 6 & 1 & 1070 \\

\end{tabular}
\caption{The number of elections with a given number of candidates $n$ and number of seats $S$ in our dataset.}
\label{table:Scottish_data_summary}
\end{table}

\section{Ballot Exhaustion}\label{section:exhaustion}
In this section we provide definitions for different types of ballot exhaustion and we describe why exhaustion may be undesirable. We provide context for our approach by first describing why ballot exhaustion has received attention in the single-winner literature.

\subsection{Ballot Exhaustion in the Single-Winner Setting}\label{section:exhaustion_IRV_subsection}
As outlined in the Introduction, most previous research on STV and ballot exhaustion focuses on single-winner IRV elections. In this setting, a ballot is \emph{exhausted} if the ballot has become empty by the final round \cite{BK15}. That is, the ballot ranks no candidate who survives to the round where a candidate earns a majority of the remaining votes. Ballot exhaustion occurs because voters often cast partial ballots, either voluntarily or because the jurisdiction limits the number of rankings.

Burnett and Kogan \cite{BK15} identify several reasons why ballot exhaustion is normatively problematic in single-winner IRV elections, which we summarize below. While Burnett and Kogan raise additional concerns about IRV more generally, those concerns do not bear directly on ballot exhaustion and are therefore outside the scope of this article. 

\begin{enumerate}

\item \textbf{Concern 1}: One of IRV's purported benefits is that it produces a majority winner; in practice, the election winner may not secure a majority of total votes cast because ballots are exhausted prior to the final head-to-head round of tabulation. We see this in Example \ref{example:AK_election}: the majority of all ballots is 94286 but Peltola secures the victory with 91265 votes. If we let the algorithm continue for one more round, eliminating Palin and leaving Peltola as the only remaining candidate, then Peltola would earn a slight majority of all ballots cast. This is not a normatively satisfying way to win a majority, but it is better than if she did not secure a majority as the only candidate left. As we explore below, some candidates in Scottish local elections do not reach the quota threshold even when we let the algorithm run until all losing candidates are eliminated.

\item \textbf{Concern 2}: Ballot exhaustion may affect electoral outcomes. That is, a different candidate might win if voters whose ballots were exhausted had their preferences represented in the final round. In Example \ref{example:AK_election}, if all 11262 voters who ranked only Begich had instead cast the ballot $B\succ Pa \succ Pe$ then Palin would have won the election instead of Peltola. Of course, it seems unlikely that all 11262 voters would support Palin over Peltola if asked to provide a second-ranked candidate, but it is \emph{a priori} possible that ballot exhaustion cost Palin the victory.
\item \textbf{Concern 3}: An exhausted ballot is essentially wasted, being discarded before the round in which the winner is determined. Voters whose ballots are exhausted lose influence over the outcome. Burnett and Kogan state that when using IRV, it is possible ``many more voters are likely to `waste their votes' by supporting candidates with a low probability of prevailing.'' In Example \ref{example:AK_election}, any voter who ranked only Begich on their ballot perhaps ``wasted'' their vote. 
\end{enumerate}

Burnett and Kogan note that ballot exhaustion may ``undermine the democratic legitimacy of IRV in the eyes of voters whose ballots become exhausted prior to the final round.'' If a candidate can secure a victory without a majority when using a voting method designed for majoritarian outcomes, Concerns (1) and (3) may highlight issues of democratic legitimacy. As part of our results and discussion, we address whether the concerns of Burnett and Kogan are borne out when we move from IRV to STV.


\subsection{Ballot Exhaustion in the Multiwinner Setting}\label{section:exhaustion_STV_subsection}

To the best of our knowledge, the term ``ballot exhaustion'' has not been formally defined in the multiwinner setting. Endersby and Towle \cite{ET14} discuss closely related phenomena in their analysis of Irish STV elections, focusing in particular on ``nontransferable ballots'' and votes that are ``not applied effectively to any candidate.'' However, their discussion is framed in terms of the mechanics of the transfer process rather than a formal, ballot-level definition. As a result, concepts such as non-transferability, contribution to a winning candidate, and ``wasted votes'' are used in overlapping ways. This highlights the need for a precise definition of ballot exhaustion that distinguishes between the transferability of a ballot at a given stage of the count and whether the ballot ultimately contributes to the election of a candidate. 

The most natural way to generalize the notion of exhaustion from the single-winner to the multiwinner context is simply to use the same definition. We adopt this approach, but also explore alternative notions of exhaustion that arise in the multiwinner setting. In particular, in STV elections it is natural to view some types of exhausted ballots as more normatively undesirable than others. For example, a ballot which is exhausted but the voter's first-ranked candidate wins a seat is of less concern than an exhausted ballot which ranks only losing candidates. We define three different types of exhausted ballots below, representing different levels of normative undesirability.

Before defining exhaustion, we must decide when we terminate the STV algorithm; i.e., we must decide which round of counting is the last round. We define this round as follows: either the $S$th seat is filled because the $S$th candidate reaches quota or
there are $k$ seats remaining and, after ordering the remaining candidates by vote total in descending order, the sum of the vote totals of all candidates ranked below position $k$ is strictly less than the vote total of the candidate in position $k$. For example, if two seats remain and the combined vote total of all trailing candidates is insufficient to overtake the second-place remaining candidate, we terminate the count. Note that this notion of \emph{final round} will result in fewer exhausted ballots than if we let the counting continue. The Scottish councils  continue counting until there are $k$ seats left unallocated and $k$ candidates remaining, unless $S$ candidates reach quota earlier. We adopt our stopping rule because once remaining trailing candidates are mathematically unable to overtake leading candidates, additional eliminations no longer affect seat allocation and, arguably, would artificially inflate exhaustion counts.

\begin{definition}
A ballot is \textbf{exhausted} if the ballot is empty in the final round of counting. That is, a ballot is exhausted if it does not rank any of the surviving candidates in the final round.

A ballot is \textbf{non-first-choice exhausted} if the ballot is exhausted and the candidate ranked first on the ballot is not an election winner. 

A ballot is \textbf{unrepresented exhausted} if the ballot is exhausted and none of the candidates ranked in the top $S$ rankings are in the winner set.

An election's \textbf{exhaustion rate} is the percentage of ballots which are exhausted; similarly for non-first-choice and unrepresented rates.

\end{definition}

Since the STV algorithm re-weights ballots which are used to elect a candidate, we also consider an election's \textbf{weight exhaustion rate}, which is the percentage of  ballot weight lost by the final round. For example, in Example \ref{example:AK_election} with $S=2$, when the STV algorithm terminates Begich has 61,869.19 votes, Palin has 59,763.99, and Peltola has 62,858, for a total of 184,491.18. Since there are 188,571 votes cast overall, this election's weight exhaustion rate is $1-(184491.18/188571)=2.2\%$. By contrast, the election's exhaustion rate is $23733/188571=12.6\%$ because all 23733 ballots which rank Peltola first are exhausted. Arguably, the 2.2\% rate is more relevant than 12.6\% because of how STV is designed.

As in the single-winner case, ballot exhaustion in multiwinner elections is caused by the presence of partial ballots. In Scottish elections voters are allowed to provide complete preferences, and thus partial ballots are cast voluntarily. Such ballots are very common: across all elections, 58.0\% ballots rank fewer than $S$ candidates, while only 13.2\% provide a complete ranking of the $n$ candidates (by ``complete ranking'', we mean a ballot of length $n$ or $n-1$). Figure \ref{figure:ballot_lengths} in the Appendix shows the percentage of ballots of given lengths, separated by $n$ and $S$. Regardless of the choice of $n$ and $S$, we see a clear pattern where most voters cast ballots of length $S$ or less, very few voters cast a ballot of length in $\{S+1, S+2, \dots, n-1\}$, and a small but non-trivial amount of voters cast a ballot of length $n$.   Given these distributions of ballot lengths, we expect many ballots to be exhausted in these elections.

We note that it is not clear why voters choose to cast so many short ballots. Perhaps voters find it cognitively difficult to rank many candidates, or voters are indifferent among candidates they do not rank. It is also possible that some voters truncate their ballots strategically because they are concerned about indirectly supporting candidates whom they do not truly support. The ballot data does not allow us to distinguish among these reasons.

Endersby and Towle \cite{ET14} highlight a set of concerns related to those of Burnett and Kogan \cite{BK15} in their analysis of Irish STV elections, focusing on the prevalence of nontransferable ballots. They emphasize that a nontrivial number of ballots are ultimately ``not applied effectively to any candidate'' because they cannot be transferred to a continuing candidate. As a result, these ballots fail to contribute to the election of any representative, raising concerns about the extent to which STV succeeds in reducing ``wasted votes.''  Moreover, Endersby and Towle find that the incidence of nontransferable ballots increases in later rounds of counting, precisely when the system relies most heavily on lower-ranked preferences. This suggests a tension between a theoretical promise of STV, that voters can express rich preference information that will be used through transfers, and the empirical reality that many voters do not provide sufficiently complete rankings. This concern implies that exhaustion may affect the electoral outcome. In addition, the complexity of the transfer process may make it unclear to voters how their ballots influence the outcome. 


To conclude this section, we note that Endersby and Towle \cite{ET14} analyze Irish STV elections using published votes-by-round tables rather than ballot-level data, limiting their ability to track individual ballots through the count. Moreover, the Irish implementation of STV uses whole-ballot surplus transfers rather than fractional reweighting.\footnote{For a complete description of the Irish STV rules, see Section 121 of the 1992 Electoral Act at \url{https://www.irishstatutebook.ie/eli/1992/act/23/section/121/enacted/en/html}. Accessed 5/5/2026.} In this setting, it is natural to focus on aggregate notions such as nontransferable votes. By contrast, the availability of ballot-level preference profile data in Scotland, together with fractional transfer rules, allows us to define and analyze exhaustion at the level of individual ballots.

\section{Results}\label{section:results}

In this section we present the empirical results of our analysis of Scottish STV elections. We begin by examining rates of ballot exhaustion and weight exhaustion, then investigate whether exhaustion can affect electoral outcomes. Finally, we analyze how often winners fail to achieve quota.

\begin{figure}
\includegraphics[width=\textwidth]{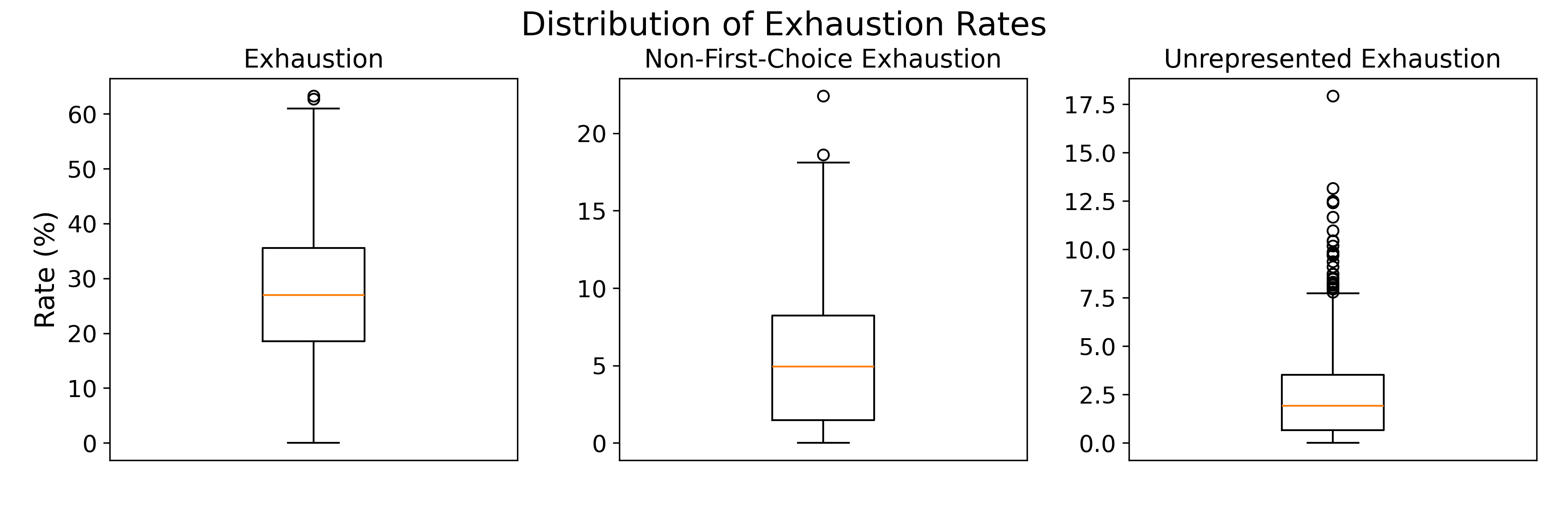}
\caption{Box plots showing the three ballot exhaustion rates across all elections in the Scottish dataset.}
\label{figure:exhaustion_rates_boxplots}
\end{figure}

\subsection{Exhaustion rates and non-transferable ballots}\label{section:results_rates_subsection}
Our top-line finding is that across all elections, 27.9\% of all ballots are exhausted, 5.4\% of ballots are non-first-choice exhausted, and 2.4\% of ballots are unrepresented exhausted. A rate of 27.9\% is quite high, but the overall weight exhaustion rate is  7.1\%. Thus, while a large number of ballots are exhausted, the majority of them are exhausted after being applied to the election of a candidate, making their exhaustion of less normative concern. Of the exhausted ballots, 19.5\% are non-first-choice exhausted and 8.5\% are unrepresented exhausted, underscoring that the large majority of exhausted ballots correspond to voters who achieve some form of representation.

Figure \ref{figure:exhaustion_rates_boxplots} shows the three types of exhaustion rates across all elections. The left plot shows that some elections have very high exhaustion rates; six elections have a rate over 60\%, for example. The middle and right plots show that many of the exhausted ballots rank a winner either in the first rank or in the top $S$ ranks, as the rates drop substantially from the first plot. In particular, the median rate of unrepresented exhaustion is less than 2.5\%, much lower than the 27.0\% median rate for exhaustion. These plots reinforce our finding that while many ballots are exhausted, many of these rank a winner in the first rank or in the top $S$ rankings, making their exhaustion of less concern. 

The left image of Figure \ref{figure:rounds_and_weight} in the Appendix shows the weight exhaustion rates across all elections. This box plot reinforces the finding that many exhausted ballots have contributed to the election of a candidate before they are exhausted, and thus the ballot weight not transferred is much lower than the raw number of ballots not transferred.

The right image of Figure \ref{figure:rounds_and_weight} plots an election's exhaustion rate versus the number of rounds the election runs. Unsurprisingly, as the number of rounds increases the exhaustion rate also tends to increase, presumably because ballots have ``more opportunity'' to exhaust. To visualize the rounds in which exhaustion tends to occur, Figures \ref{figure:exhaustion_by_round}-\ref{figure:exhaustion_by_round_unrepresnted} in the Appendix illustrate rounds in which ballots are exhausted, for elections which last $k$ rounds. For a number of rounds $k$, Figure \ref{figure:exhaustion_by_round} shows the percentage of all ballots exhausted across all elections at the end of round $k$, for elections which run at least $k$ rounds. The figure shows the percentage of total ballots that are exhausted when moving from round $k-1$ to $k$, and shows the cumulative percentage of ballots exhausted by round $k$.

For example, suppose $k=4$. There are 915 elections in our dataset for which STV runs at least 4 rounds, and these elections contain 4,823,750 ballots in total. Of these ballots, 185,575 are exhausted in moving from round 3 to round 4, accounting for $185575/4823750=3.85\%$ of all ballots. Thus, the blue plot has a point at $(4,3.85\%)$. The orange curve provides a cumulative percentage for rounds 1 through $k$. Across the 915 elections which last at least 4 rounds, 613059 ballots are exhausted by round 4, giving a cumulative percentage of 12.70\%. Thus, the orange curve has a point at $(4,12.70\%)$. We note the cumulative curve is not necessarily monotonic because elections with different round counts contribute to different values of $k$, causing the ballot denominator to change for different values of $k$. The plots for non-first-choice and unrepresented exhaustion (Figures \ref{figure:exhaustion_by_round_non_first_choice} and \ref{figure:exhaustion_by_round_unrepresnted}) are constructed similarly. 


Figure \ref{figure:exhaustion_round_by_round} in the Appendix shows the percent of ballots not transferred and the amount of weight not transferred from round $(k-1)$ to round $k$ for elections which last at least $k$ rounds. In contrast to the previous figures, the denominator in these figures is not the total amount of ballots, but is instead the number of ballots which could potentially transfer in that round (blue) or the amount of ballot weight that could transfer in that round (orange). For example, suppose $k=4$. Of all elections which last at least 4 rounds, there is a total of 591755 ballots which are in the ``transfer pile'' at the end of round 3; i.e., there are 591755 ballots for which the candidate elected or eliminated in round 3 is currently ranked first on the ballot. Of these ballots, 185575 currently rank only one candidate at the end of round 3, and thus do not transfer to round 4. Thus, $185575/591755=31.36\%$ of ballots which currently rank the  candidate elected or eliminated in round 3 do not transfer to round 4, and the blue curve has a point at $(4, 31.36\%)$. The orange curve is constructed similarly, but measures the percent of round-by-round weight which does not transfer. For example, of the elections which last at least 4 rounds, the total weight which could transfer after round 3 is 165625.25, of which 40246.92 does not transfer, giving a point of $(4, 24.30\%)$ on the orange curve.

Figures \ref{figure:exhaustion_by_round}-\ref{figure:exhaustion_round_by_round} all show that the amount of exhaustion increases in later rounds, an unsurprising result. Figure \ref{figure:exhaustion_round_by_round} shows that in later rounds at least 40\% of ballots or weight are lost from round-to-round, for example.  

\begin{figure}[h]
\includegraphics[width=\textwidth]{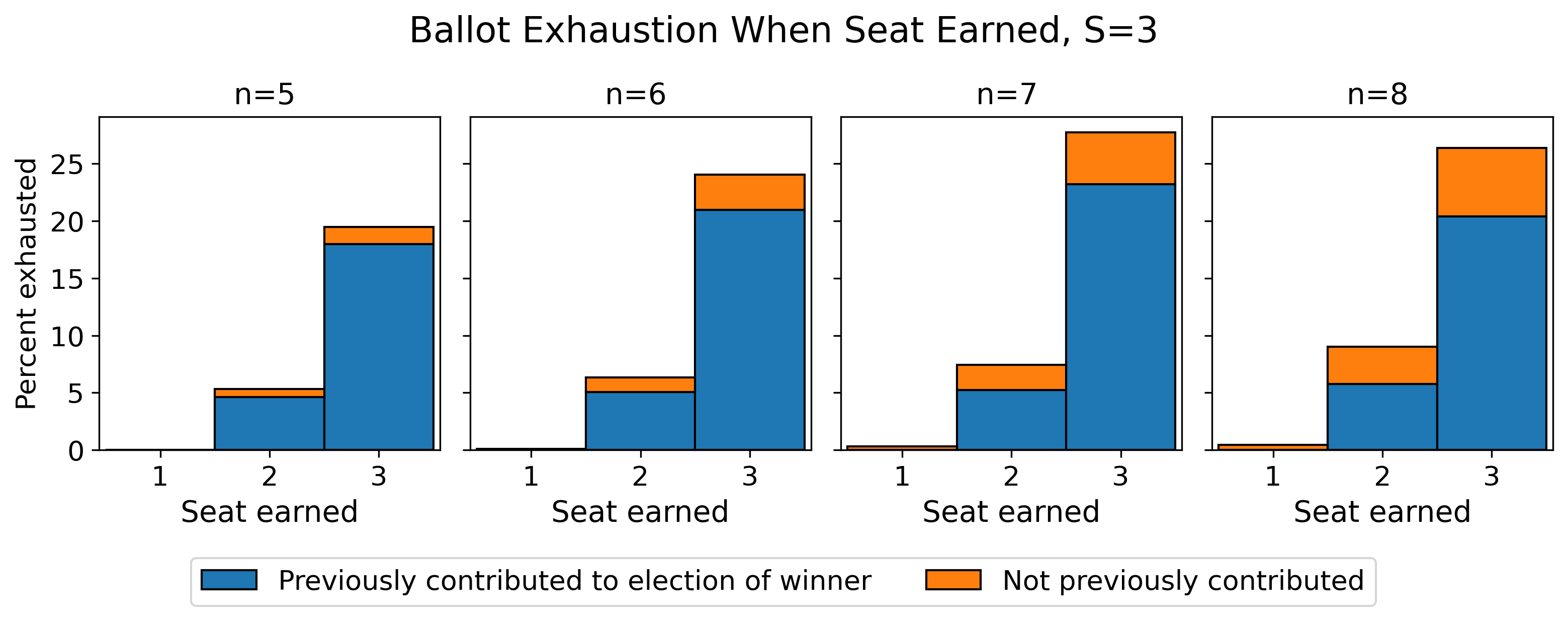}
\caption{The percentage of ballots exhausted grouped by when a seat is earned in 3-seat elections. The height of each bar gives the percentage of ballots exhausted when the $k$th seat is earned, and the orange shows the portion of ballots exhausted without being applied to the vote count of a winner when they won their seat.}
\label{figure:seat_by_seat_histograms_3_seats}
\end{figure}

\begin{figure}[h]
\includegraphics[width=\textwidth]{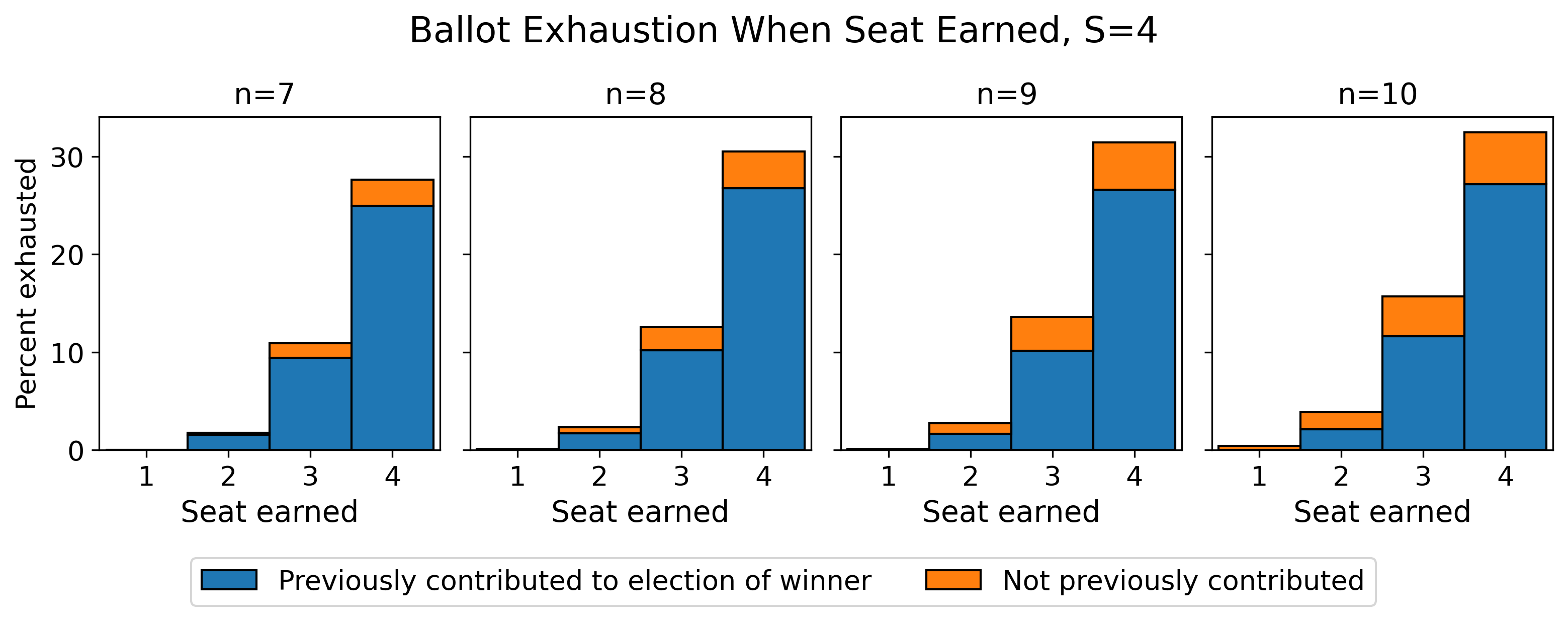}
\caption{The percentage of ballots exhausted grouped by when a seat is earned in 4-seat elections. The height of each bar gives the percentage of ballots exhausted when the $k$th seat is earned, and the orange shows the portion of ballots exhausted without being applied to the vote count of a winner when they won their seat.}
\label{figure:seat_by_seat_histograms_4_seats}
\end{figure}

Figures \ref{figure:seat_by_seat_histograms_3_seats} and \ref{figure:seat_by_seat_histograms_4_seats} further clarify the nature of exhaustion by grouping exhausted ballots according to when a seat is earned, rather than grouping strictly by round. The figures provide histograms for the eight choices of $(n,S)$ with the most elections, and display the percentage of ballots exhausted in the round where the $k$th seat is earned. Across both 3-seat and 4-seat elections, the large majority of ballots exhausted when later seats are filled have already contributed to the election of a winner earlier in the count, where by ``contribute'' we mean that the ballot  ranked a winning candidate in first when that candidate won their seat. Most exhaustion occurs when the final seat is earned, and much of this exhaustion corresponds to ballots that previously contributed to the election of another candidate. This reinforces our earlier conclusion that raw exhaustion rates overstate the extent to which voters fail to obtain representation, since many exhausted ballots have already played a meaningful role in determining the winner set.

\subsection{Does exhaustion affect electoral outcomes?}\label{section:results_electoral_outcomes_subsection}

This question is difficult to answer because we do not know how a voter would fill in blank rankings if the voter had decided to provide a complete ranking. If all voters who cast partial ballots provided a complete ballot instead, it is \emph{a priori} possible that winner sets might significantly differ from what we see when using the actual ballots. To investigate the potential effects of exhaustion on electoral outcomes, we ran three experiments, two of which are ``worst-case'' and one which is (hopefully) more realistic.

In our first experiment, for each election we added the $(S+1)$st ranked candidate (the candidate who had the most votes among losing candidates in the final round), denoted $L_1$, to the first empty ranking of all ballots which did not already rank this candidate. It is possible, although of course highly unlikely, that all voters who cast partial ballots not containing $L_1$ would have ranked $L_1$ next on their ballot if asked to rank one additional candidate. This model represents an extreme  scenario in which the ``strongest loser''  receives maximal additional support from all partial ballots which did not already rank them. Under this model, $L_1$ becomes a winner in 745 of the 1070 elections, meaning that it is theoretically possible ballot exhaustion had an effect on the outcome of 69.6\% of the elections. In each of the 745 elections the only change to the winner set is $L_1$ replaces one candidate, meaning that up to $745/3718=20.0\%$ of available seats could potentially have been affected by ballot exhaustion.

In our second experiment, for each election we found the $(S+1)$st ranked candidate and the $(S+2)$nd ranked candidate, denoted $L_2$, and added these candidates to the first available skipped rankings on a ballot as follows. If the ballot already ranked $L_1$ but not $L_2$ then we added $L_2$ to the first skipped ranking on the ballot, and vice versa. If the ballot ranked neither candidate we placed them in the first two skipped rankings in random order. Since this experiment involves random choice, we ran it several times. The maximum number of elections in which we saw a change to the winner set was 754, representing 70.5\% of all elections. Of these, $L_1$ gained a seat but not $L_2$ in 650, $L_2$ gained a seat but not $L_1$ in 78, and both gained a seat in 26, meaning that overall in this scenario $780/3718=21.0\%$ of available seats could have been affected by exhaustion. As in the first experiment, these changes overwhelmingly consist of single-seat substitutions, indicating that even under very extreme ballot completions, exhaustion does not produce large swings to a given winner set. 

These two experiments suggest that exhaustion could significantly affect electoral outcomes  in a worst-case, theoretical sense. Our third experiment adds candidates to skipped rankings in a more realistic way using a proportional ballot-completion model. Intuitively, this approach assumes that voters who leave rankings blank would be more likely to rank candidates in proportion to their overall levels of support, rather than systematically favoring specific losing candidates. To be more precise, in this model  partial ballots are extended iteratively using the observed behavior of voters who cast longer ballots with the same prefix. For example, suppose a ballot ranks only candidate $A$. We look at all ballots beginning with $A$ that rank at least one additional candidate and use their second-choice distribution to extend the $A$-only ballots proportionally. Thus, if longer ballots beginning with $A$ next rank $B$, $C$, and $D$ in proportions $10:20:25$, then the $A$-only ballots are split among $A\succ B$, $A\succ C$, and $A\succ D$ in those same proportions, up to integer rounding. We then repeat this process for ballots of length two, then length three, and so on. We continue  until all ballots are complete, or we run out of ways to continue adding candidates to a ballot. For example, suppose $n=5$ and we want to extend a ballot like $A\succ B \succ C$ to add a fourth candidate. If there is no ballot longer than three beginning with $A\succ B \succ C$ in the ballot data then we do not extend the ballot further.

Of course, we cannot expect any model to completely capture unexpressed voter preferences. This proportional model represents a reasonable attempt, conditioned on observed preferences. A more sophisticated model could incorporate data such as the precinct in which a ballot was cast, but Scottish councils do not provide such information.

Under this proportional model of ballot completion, the winner set changes in only 130 elections, a much smaller number than what was observed in the two worst-case experiments. For each of these elections only one of the original winners loses a seat, meaning that in this more realistic setting ballot exhaustion potentially affects $130/3718=3.5\%$ of the available seats. Moreover, in 11 of these elections ballot completion simply swaps a candidate of one party for a candidate of the same party\footnote{Scottish elections sometimes have independent candidates; we count each independent candidate as belonging to their own individual party, meaning these 11 elections do not include outcomes where one Independent is swapped for another.}, meaning that only 119 (3.1\%)  seats are meaningfully affected by exhaustion.

\subsection{How often do STV winners fail to reach quota?}\label{section:results_quota_failure_subsection}

Burnett and Kogan \cite{BK15} point out that ballot exhaustion can allow an IRV winner to secure their victory without earning a majority of all votes cast (Concern 1). Graham-Squire and McCune \cite{GSM25} show this occurs with non-trivial frequency in American IRV elections. In elections without an initial majority winner, approximately 52\% produce a final IRV winner who does not obtain a majority of all ballots cast when the IRV algorithm terminates. This number may be inflated because some American jurisdictions which use IRV limit the number of candidates ranked by voters. For example, in Minneapolis a voter can rank only three candidates, regardless of how many candidates are running. Scottish elections do not have this restriction. 

Of the 3718 seats available across the Scottish dataset, 28.0\% are earned by winners who have not achieved quota when our STV algorithm terminates. If we allow the algorithm to run longer, until we reach an elimination round where there are $k$ seats left to be filled and $k+1$ surviving candidates, the percentage drops to 25.0\%, which is still quite large.  If we let the algorithm run until there are $k$ seats left to be filled and $k$ surviving candidates, 3.4\% of the seats are filled by candidates who do not achieve quota. That is, 127 of 3718 seats are filled by candidates who cannot reach quota even after all losers are eliminated. Thus, ballot exhaustion could produce ``legitimacy'' challenges, where winners are sometimes elected without crossing the quota threshold, even after eliminating all losing candidates.

Table  \ref{table:Dumgal_example} shows the votes-by-round table for the 2017 election in the Stranraer and the Rhins Ward of the Dumfries and Galloway council area. This examples illustrates how such quota failures can affect multiple candidates. Our algorithm would terminate at Round 6, since there are two seats remaining and it is clear those  seats are won by Sloan and Surtees. Since the quota is 1055, two of the four seats are won by candidates who do not reach quota. Even if we let the algorithm run an additional round, neither Sloan nor Surtees can reach 1055 votes. Thus, this example represents a worst-case outcome from a ``democratic legitimacy'' standpoint, where two candidates cannot achieve quota even if they are the only two candidates remaining.

\begin{table}

\begin{tabular}{lccccccc}
\multicolumn{8}{c}{Quota $= 1055$, $S=4$}\\
\hline
Candidate & Round 1 & Round 2 & Round 3 & Round 4 & Round 5 & Round 6 & Round 7 \\
\hline
Willie Scobie & \textbf{1925} &  &  &  &  &  &  \\
Andrew Giusti & \textbf{1703} &  &  &  &  &  &  \\
Ros Surtees & 765 & 869.40 & 884.62 & 903.35 & 919.09 & 953.34 & \textit{1013.39} \\
Tommy Sloan & 312 & 589.94 & 661.86 & 694.11 & 723.69 & 774.28 & \textit{912.14} \\
Tracy Davidson & 181 & 251.95 & 317.40 & 338.72 & 381.51 &  &  \\
Marion McCutcheon & 166 & 259.55 & 332.23 & 363.55 & 418.26 & 533.25 &  \\
Robert McCrae & 123 & 164.13 & 231.48 & 242.73 &  &  &  \\
Chris Collings & 95 & 134.77 & 174.72 &  &  &  &  \\
\hline
\end{tabular}

\caption{The votes-by-round table for the 2017 election in the Stranraer and the Rhins Ward of the Dumfries and Galloway council area. Two of the four seats are earned by candidates who do not reach quota.}
\label{table:Dumgal_example}
\end{table}

\subsection{Robustness check: Australian elections}

To examine the extent to which our results depend on the Scottish context, we repeat our main analysis for 477 local elections from New South Wales, Australia. The ballot data for more than 477 elections is available, but to more directly compare to the Scottish results we keep only elections with 2, 3, 4, or 5 seats, where the number of candidates is at most 15. The Australian data is available at \url{https://vote.andrewconway.org}.

The NSW elections provide a useful comparison to the Scottish elections because the two implementations of STV are broadly similar but differ in two important respects. First, NSW requires voters to rank a minimum number of candidates, equal to half the number of seats available, rounded up, whereas Scottish voters may rank as few as one candidate. Second, unlike the Scottish procedure, the NSW implementation transfers only integer vote totals, with fractional remainders disregarded. For a complete description of the NSW rules, see \url{https://elections.nsw.gov.au/getmedia/7caa2c43-a82a-4da8-a8f4-523ef647057b/el-3935-functional-requirements-for-lg-pr-vote-count.pdf}.

Table~\ref{tab:scotland_nsw} compares exhaustion rates across the two datasets. The overall exhaustion rate is noticeably lower in NSW than in Scotland, at 20.9\% compared to 27.9\%. This difference may partially reflect the NSW minimum-ranking requirement. The differences are smaller for our more restrictive measures of exhaustion: non-first-choice exhaustion is 3.9\% in NSW compared to 5.4\% in Scotland, while unrepresented exhaustion is somewhat higher in NSW, at 3.2\% compared to 2.4\%.

Importantly, the weight exhaustion rates are quite similar: 6.6\% in NSW and 7.1\% in Scotland. Thus, despite differences in electoral context, ballot requirements, and the implementation of STV, our primary finding is robust across the two datasets. Raw ballot exhaustion substantially overstates the amount of voting weight lost during the STV counting process.

\begin{table}[ht]
\centering

\begin{tabular}{lrr}
\hline
\textbf{Measure} & \textbf{Scotland} & \textbf{NSW Australia} \\
\hline
Elections                    & 1,070      & 477       \\
Ballots                      & 5,483,680  & 6,588,755 \\
Regular exhaustion           & 27.9\%     & 20.9\%    \\
Non-first-choice exhaustion  & 5.4\%      & 3.9\%     \\
Unrepresented exhaustion     & 2.4\%      & 3.2\%     \\
Weight exhaustion            & 7.1\%      & 6.6\%     \\
\hline
\vspace{.001 in}
\end{tabular}
 \caption{A comparison of exhaustion rates in the Scottish and Australian datasets.}
\label{tab:scotland_nsw}
\end{table}

\section{Discussion}\label{section:discussion}
In this section we discuss our results by revisiting the concerns about ballot exhaustion outlined in Section \ref{section:exhaustion} and analyzing rates of rejected ballots. We also briefly explore an alternative notion of ballot exhaustion through ballot re-weighting.

\subsection{Concerns about Ballot Exhaustion}
Our results in the previous section allow us to determine if the concerns of Burnett and Kogan \cite{BK15} and Endersby and Towle \cite{ET14} (Sections \ref{section:exhaustion_IRV_subsection} and \ref{section:exhaustion_STV_subsection}) apply to the Scottish context. Concern 1 of Burnett and Kogan translates to our multiwinner setting. As outlined in Section \ref{section:results_quota_failure_subsection}, many candidates fail to achieve quota when we let the algorithm until there is only one losing candidate left. Furthermore, if we let the algorithm run until there are no losing candidates left, a non-trivial number of winners still fail to reach the quota threshold.

On the other hand, Concern 2 does not translate. When we make a reasonable guess about how voters would fill in blank rankings, we see that election outcomes are not markedly different (Section \ref{section:results_electoral_outcomes_subsection}). These results suggest that ballot exhaustion does not significantly affect which candidates earn seats.

Concern 3 also does not meaningfully translate to our multiwinner setting. The large majority of exhausted ballots correspond to voters who achieve some form of representation; either their top-ranked candidate or a candidate ranked in their top $S$ rankings gets a seat on the council. That is, most voters whose ballot is exhausted have their ballot contribute to an election of a candidate, and thus such ballots are not ``wasted'' in any meaningful sense.

In response to concerns about ballot exhaustion, one possible argument is that voters whose ballots become exhausted are simply indifferent among the remaining candidates, so the loss of influence is not especially troubling. For example, in the Alaska IRV election in Example \ref{example:AK_election}, it is possible that all voters who cast a Begich-only ballot are truly indifferent between Palin and Peltola. In the single-winner case, Burnett and Kogan \cite{BK15} reject this argument, stating ``in each of the cases we examine... the ideological differences between candidates in the final runoff were stark, making it highly implausible that indifference explains the high rate of incomplete ballots.'' While this claim is persuasive for the four elections they study, and may apply to IRV elections more generally, it is less convincing in the multiwinner setting examined here. Of the 50 elections with the highest exhaustion rates, 35 exhibit a similar dynamic: the election contains two candidates from the Scottish National Party (SNP), and the STV algorithm terminates in a round where these two candidates compete for one remaining seat. In such cases, it is quite plausible that many voters are indifferent about which SNP candidate joins the council, especially if these voters have already achieved representation.  An additional 7 of the 50 display the same dynamic where two ideologically similar candidates (but not from the SNP) compete for the last seat. While Burnett and Kogan's argument may be persuasive in the single-winner context, it does not translate to our multiwinner setting.

Furthermore, it is conceivable that the ideological differences between the surviving candidates in the last round are stark, yet many voters are still indifferent between them. For example, if the two remaining candidates are perceived as an extreme left and an extreme right candidate, many centrist voters may be indifferent about which extreme candidate joins the council, especially if at least one centrist has earned a seat. For example, in the election with the 40th largest exhaustion rate, the last two surviving candidates are a Conservative (generally considered the rightmost party in Scotland) and a member of the Green Party (generally considered a far-left party in Scotland). Thus, it is plausible a substantial portion of the voters ``in the middle'' are indifferent between the two candidates. 

In the large majority of the 50 elections with the highest exhaustion rates, many voters may perceive relatively little difference in representational value between the candidates who survive to the final round. In contrast to the analysis of Burnett and Kogan, voter indifference is a reasonable (though not proven) explanation for high exhaustion rates in our dataset.

Finally, Endersby and Towle are concerned that the complexity of the transfer process makes opaque to a voter how their ballot influenced the election's outcome. There is no  good solution to this issue in jurisdictions which do not make digital copies of their ballot data,  which is still the case for Ireland. However, this problem is solvable for Scottish councils. Election officials could easily add a website tool that allows voters to enter a ballot and view its path through the election. An example of such output is provided in Table \ref{table:ballot_flow}, which shows how a sample ballot flows through the transfer process for the election displayed in Table \ref{table:Dumgal_example}. We do not argue this is the optimal way to display such information, but this concern of Endersby and Towle is easily addressed with modern computational tools and a good web designer, in jurisdictions which use fractional transfers and produce ballot record files.

\begin{table}

\begin{tabular}{c|c|c|c}
\multicolumn{4}{l}{\textbf{Ballot: Giusti $\succ$ McCrae $\succ$ Sloan $\succ$ Collings } }\\
\hline
\hline
    Round&Current Ballot&Ballot Weight&Contribution\\
    \hline
    1 & Giusti $\succ$ McCrae $\succ$ Sloan $\succ$ Collings & 1 & $-$\\
    2 & Giusti $\succ$ McCrae $\succ$ Sloan $\succ$ Collings & 1 & 1 Vote to Election of Giusti; \\ &&&Giusti keeps 0.62\\
    3 & McCrae $\succ$ Sloan $\succ$ Collings & 0.38 & $-$\\
    4 & McCrae $\succ$ Sloan & 0.38 & $-$\\
    5 & Sloan & 0.38 & $-$\\
    6 & Sloan & 0.38 & 0.38 Votes to Election of Sloan\\
\end{tabular}
\vspace{.1 in}
\caption{An example of displaying the``ballot flow'' of a specific ballot through the STV election from Table \ref{table:Dumgal_example}.}
\label{table:ballot_flow}
\end{table}

\subsection{Rejected ballots}

As Endersby and Towle \cite{ET14} observe, the term ``wasted vote'' could be interpreted broadly to include ballots which are ``exhausted'' because they are rejected before the first round of counting. In Scotland, ballots are rejected for a variety of reasons, including  that they do not clearly indicate a voter's first-ranked candidate.  Our main analysis does not include rejected ballots, but they warrant some investigation because arguably they are the most problematic kind of exhausted ballots.


Every council reports the number of rejected ballots in each ward election, and thus we have access to each election's \emph{rejected ballot rate}, which is the percentage of cast ballots which are rejected. Figure \ref{figure:rejected_ballots} in the Appendix shows the rejection rates across all elections, separated by year. The median rejection rates are approximately 1.8\% for each year, and the rates do not tend to decrease over time in the aggregate. Proponents of IRV and STV often argue that ballot error rates should decrease over time as voters become accustomed to the voting method; we do not see this in Scotland. The reason may be that STV is used only for local elections, which occur once every five years. If Scottish voters used the system more often and for a wider variety of elections, perhaps the rate would decrease over time.

Across all elections, approximately 1.9\% of all ballots cast are rejected. This rate is high compared to Irish STV elections, where Endersby and Towle report a rate of 1.0\% for their dataset. It is not clear what ballot rejection rate would be acceptable but 1.9\% seems too high, a sentiment echoed by the UK Electoral Commission.\footnote{See the Commission's statement after the 2022 election cycle at \url{https://www.electoralcommission.org.uk/media-centre/scottish-council-elections-well-run-still-too-many-votes-discounted}. Accessed 5/4/2026.} Even if all concerns of Burnett and Kogan were addressed, the rejection rate is problematic.

\subsection{Effective Exhaustion Through Re-Weighting} \label{section:results_effective_exhaustion_subsection}
Our definitions of exhaustion focus on whether a ballot remains ``active,'' but from the perspective of voter influence, a ballot with zero weight is functionally indistinguishable from an exhausted ballot. Due to the mechanics of the STV algorithm, it is possible for a ballot to be active in the sense that it still ranks candidates, but the weight of the ballot is zero. For example, if a voter ranks candidate $A$ first and  $A$'s initial number of first-place votes equals the quota, then after $A$ is elected this voter's ballot has weight zero for all subsequent rounds. Even if the voter ranks more candidates, the voter has no further influence on the election. This outcome can  be interpreted as a form of ``effective'' ballot exhaustion.

Across all elections, 6273 ballots are not exhausted but have weight zero by the final round, thereby being ``effectively exhausted'' despite not satisfying Definition 1. Furthermore, 8569 (respectively 17105) ballots are not exhausted but have weight less than 0.0001 (respectively 0.001) by the final round; these ballots have such a small weight that they could also be classified as effectively exhausted. These totals represent a tiny fraction of all 5.4 million ballots, but such re-weighting can have electoral consequences.

For example, the 2022 election in the Rutherglen Central and North ward of the South Lanarkshire council area unfolded as shown in Table \ref{table:South_Lanark_ex}. The election contained two Labour candidates: Martin Lennon and Jack McGinty. Note that any ballot which ranks Lennon  first is re-weighted to zero after round 2 because Lennon achieves quota exactly, and the vote totals for the other candidates do not change in round 3 because there is no surplus. There are 988 voters who rank Lennon first and McGinty second, all of whose ballots effectively no longer count after round 2 even though they are not exhausted. This is unfortunate for McGinty, who loses by approximately 13 votes in round 5. 

\begin{table}
\begin{tabular}{lccccccc}
\multicolumn{8}{c}{Quota $= 1270$, $S=3$}\\
\hline
Candidate & Round 1 & Round 2 & Round 3 & Round 4 & Round 5 & Round 6 & Round 7 \\
\hline
Martin Lennon & 1245 & \textbf{1270} &  &  &  &  &  \\
Janine Calikes & 1188 & 1265 & 1265 & \textbf{1338} &  &  &  \\
Andrea Cowan & 725 & 796 & 796 & 840 & 899.56 & 1029.22 & \textit{1163.27} \\
Libby Fox & 609 & 612 & 612 & 730 & 730.41 & 881.46 &  \\
Jack McGinty & 531 & 567 & 567 & 716 & 717.88 &  &  \\
Gloria Adebo & 517 & 544 & 544 &  &  &  &  \\
Alex McRae & 261 &  &  &  &  &  &  \\
\hline
\end{tabular}
\caption{The votes-by-round table for the 2022 election in the Rutherglen Central and North ward of the South Lanarkshire council area.}
\label{table:South_Lanark_ex}
\end{table}

Ultimately, so few ballots are  ``effectively exhausted'' by re-weighting across our dataset that this type of exhaustion is not a serious concern. However, the existence of these types of ballots helps illustrate that sometimes candidates are elected through ``accidents of quota.'' In the example in Table \ref{table:South_Lanark_ex}, if the quota were one vote larger then McGinty would win a seat because Lennon's election would have been delayed by a round, allowing for a large transfer from Lennon to McGinty after Adebo is eliminated in round 2.

\section{Conclusion}\label{section:conclusion}

In this article we study ballot exhaustion in multiwinner STV elections using a large dataset of  Scottish local government elections. We introduce several formal notions of exhaustion in the STV setting and show that while exhausted ballots are common, most exhausted ballots have already contributed to the election of a candidate when they become inactive. In particular, the overall ballot exhaustion rate is substantially larger than the corresponding weight exhaustion rate, indicating that raw counts of exhausted ballots overstate the amount of voter influence lost during the counting process.

We also examine whether ballot exhaustion can affect electoral outcomes. Under extreme assumptions about how voters with partial ballots might complete their rankings, exhaustion can potentially alter a substantial number of outcomes. However, under a proportional ballot-completion model based on observed voter preferences, the effect is much smaller, changing only a small fraction of seats. In addition, we show that a nontrivial number of winners fail to achieve quota, even after all losing candidates are eliminated, raising questions about how quota thresholds should be interpreted in elections with substantial exhaustion.

This study has several limitations. First, our dataset consists entirely of Scottish local government elections, and thus the results may not generalize to other countries or to elections held at different governmental levels. Second, while our ballot-completion experiments provide insight into the potential effects of exhaustion, no completion model can fully recover voters' unexpressed preferences. Finally, our analysis focuses on the Scottish implementation of STV; other forms of STV may produce different results.

 Our results suggest that ballot exhaustion in multiwinner STV elections is a more nuanced phenomenon than raw exhaustion rates alone might imply. While exhaustion is common, most exhausted ballots still contribute meaningfully to representation, and the practical effect of exhaustion on electoral outcomes appears limited in most cases.

\clearpage

\section*{Appendix: Figures}

\begin{figure}[h]
\centering
 \begin{tabular}{ccc}
\includegraphics[width=70mm]{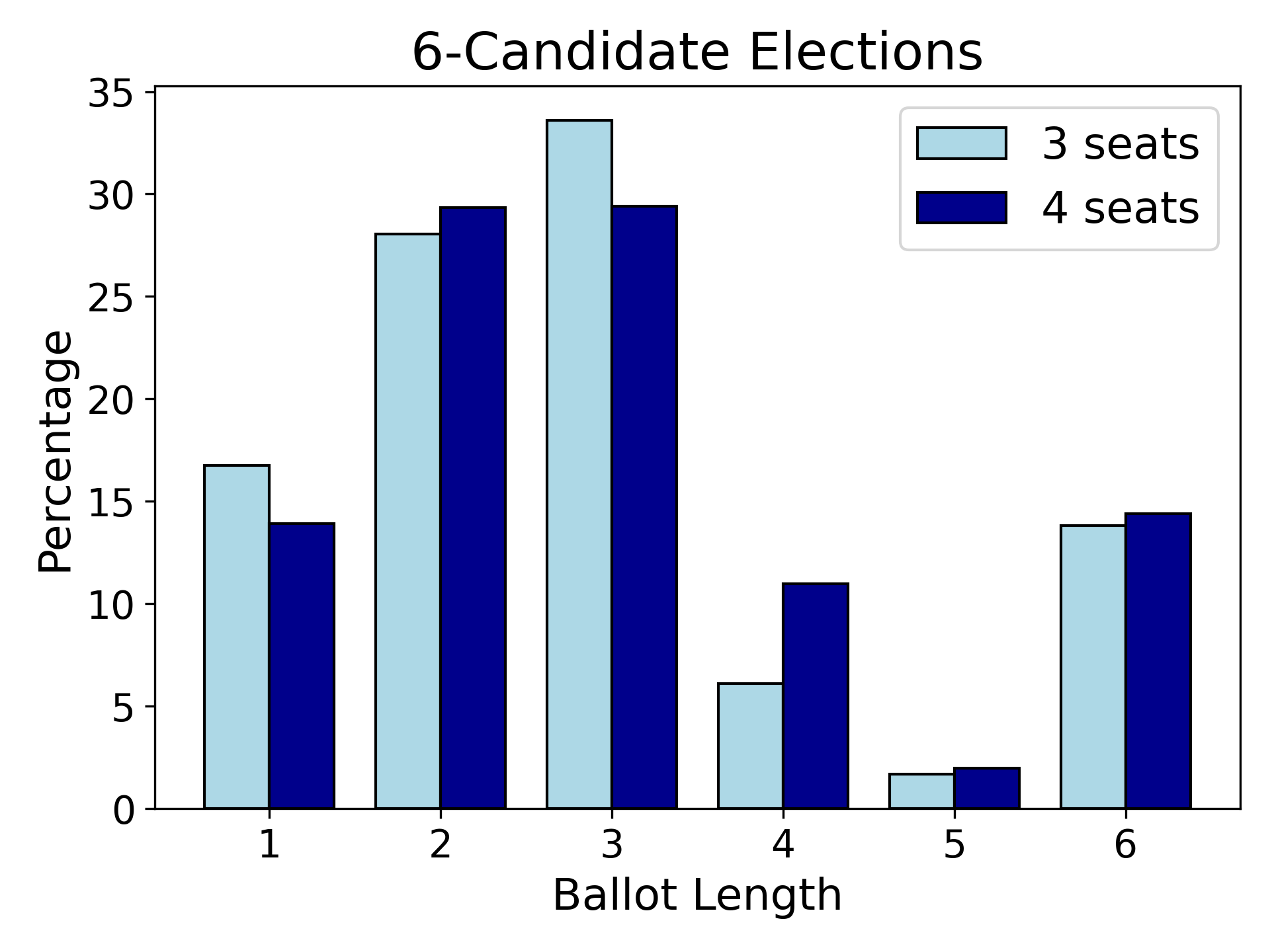}&& \includegraphics[width=70mm]{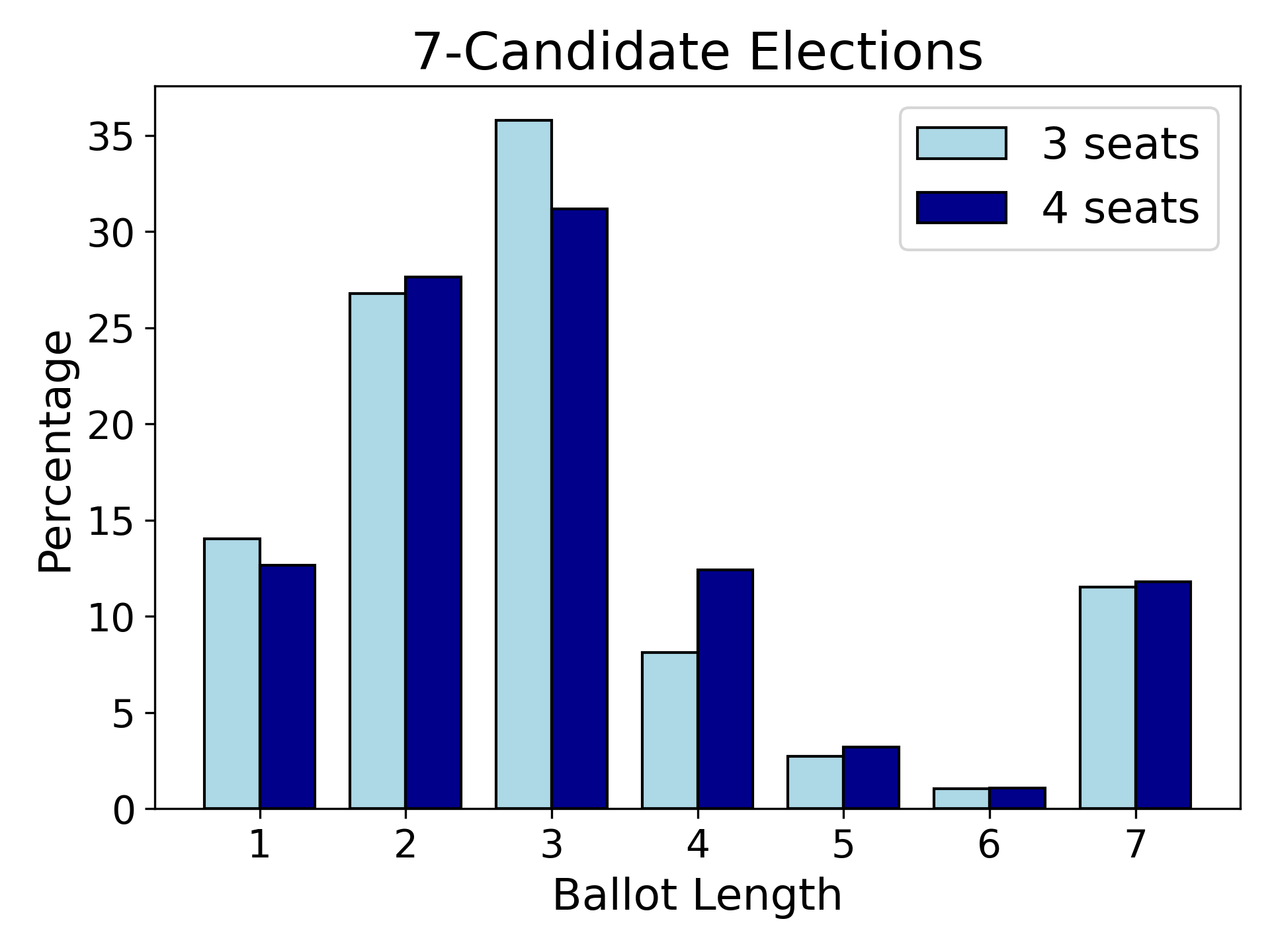}\\
\includegraphics[width=70mm]{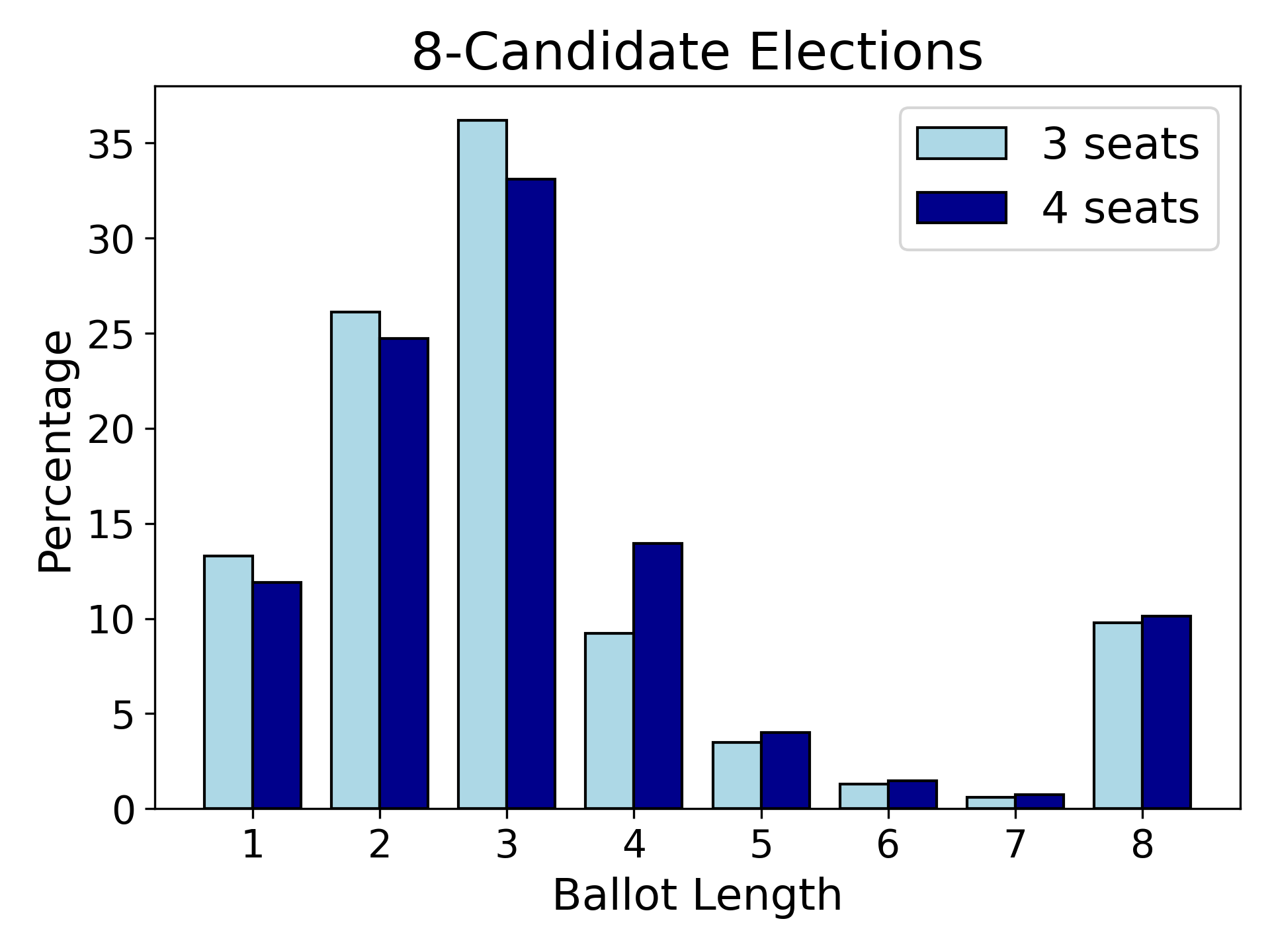}&& \includegraphics[width=70mm]{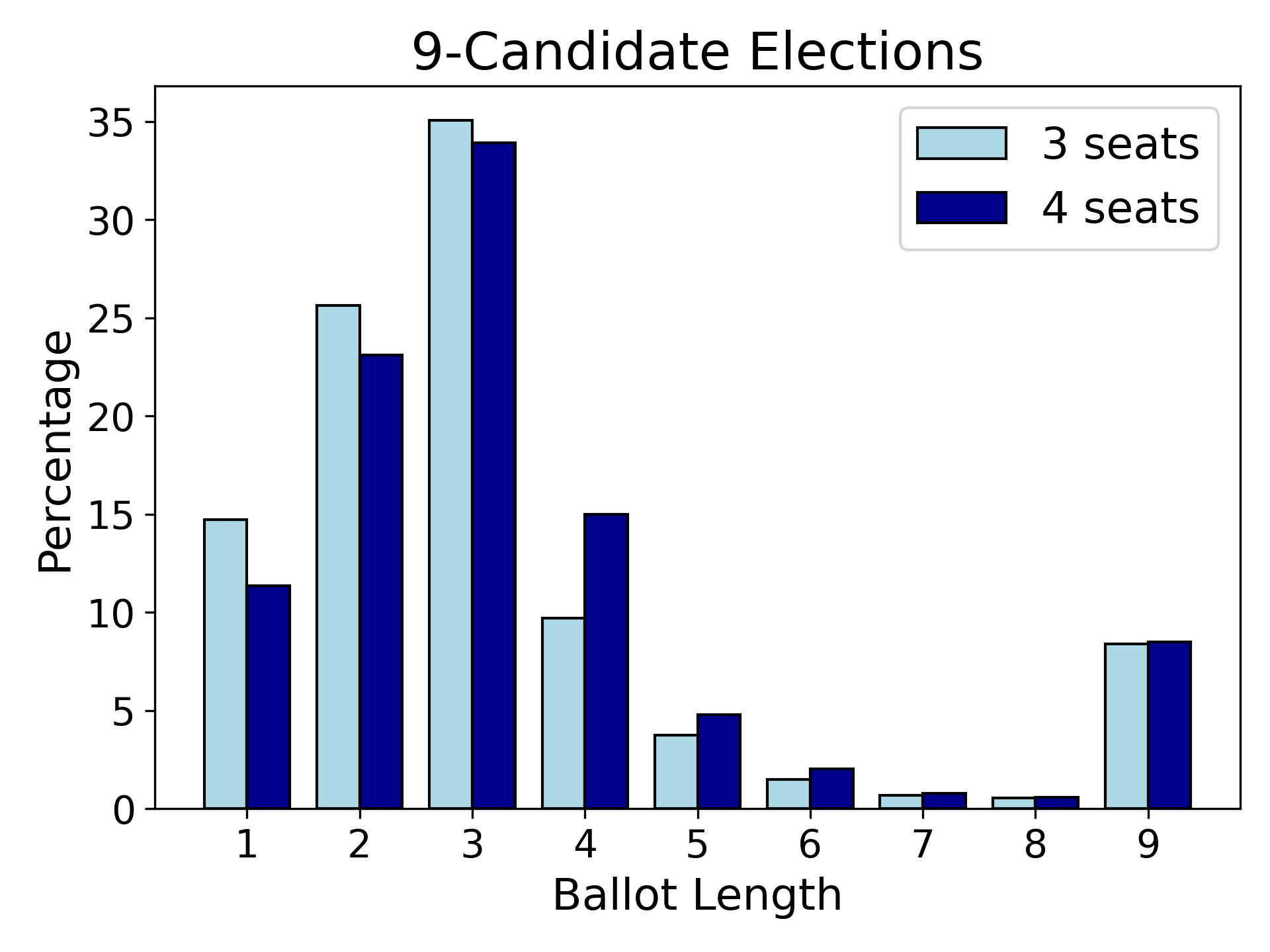}\\

 \end{tabular}
 \caption{The percentage of ballots of a given length for a choice of $n$ and $S$ where $n \in \{6,7,8,9\}$ and $S \in \{3,4\}$. For example, across all elections with $n=6$ and $S=3$, 16.8\% of ballots have length one.}
 \label{figure:ballot_lengths}
\end{figure}

\begin{figure}
\begin{tabular}{cc}
\includegraphics[width=70mm,height=61mm]{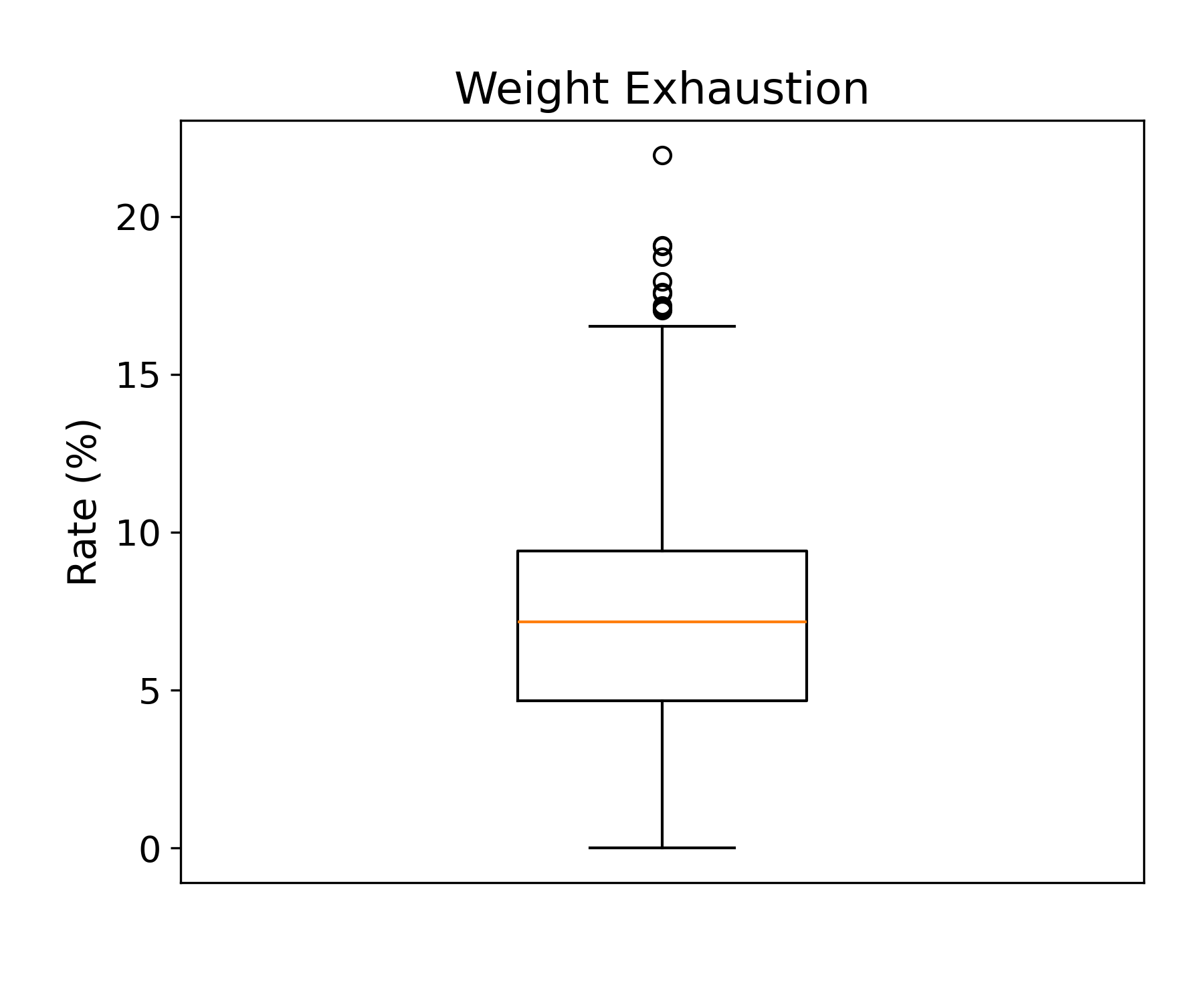}&\includegraphics[width=70mm]{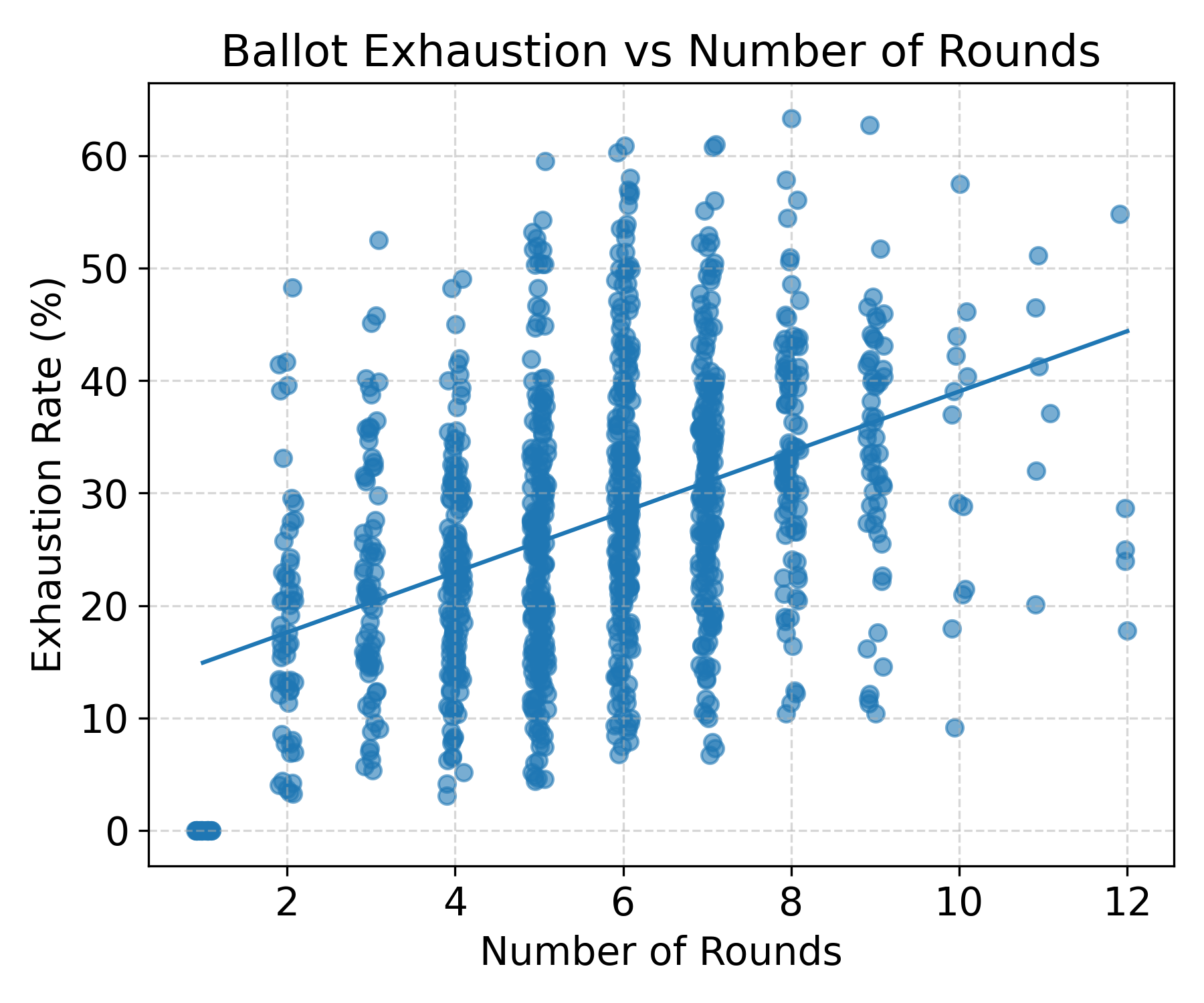}

\end{tabular}
\caption{(Left) The exhaustion rate of all elections versus the number of rounds. (Right) The weight exhaustion rates of all elections.}
\label{figure:rounds_and_weight}
\end{figure}

\begin{figure}[h]
\centering
\includegraphics[width=130mm]{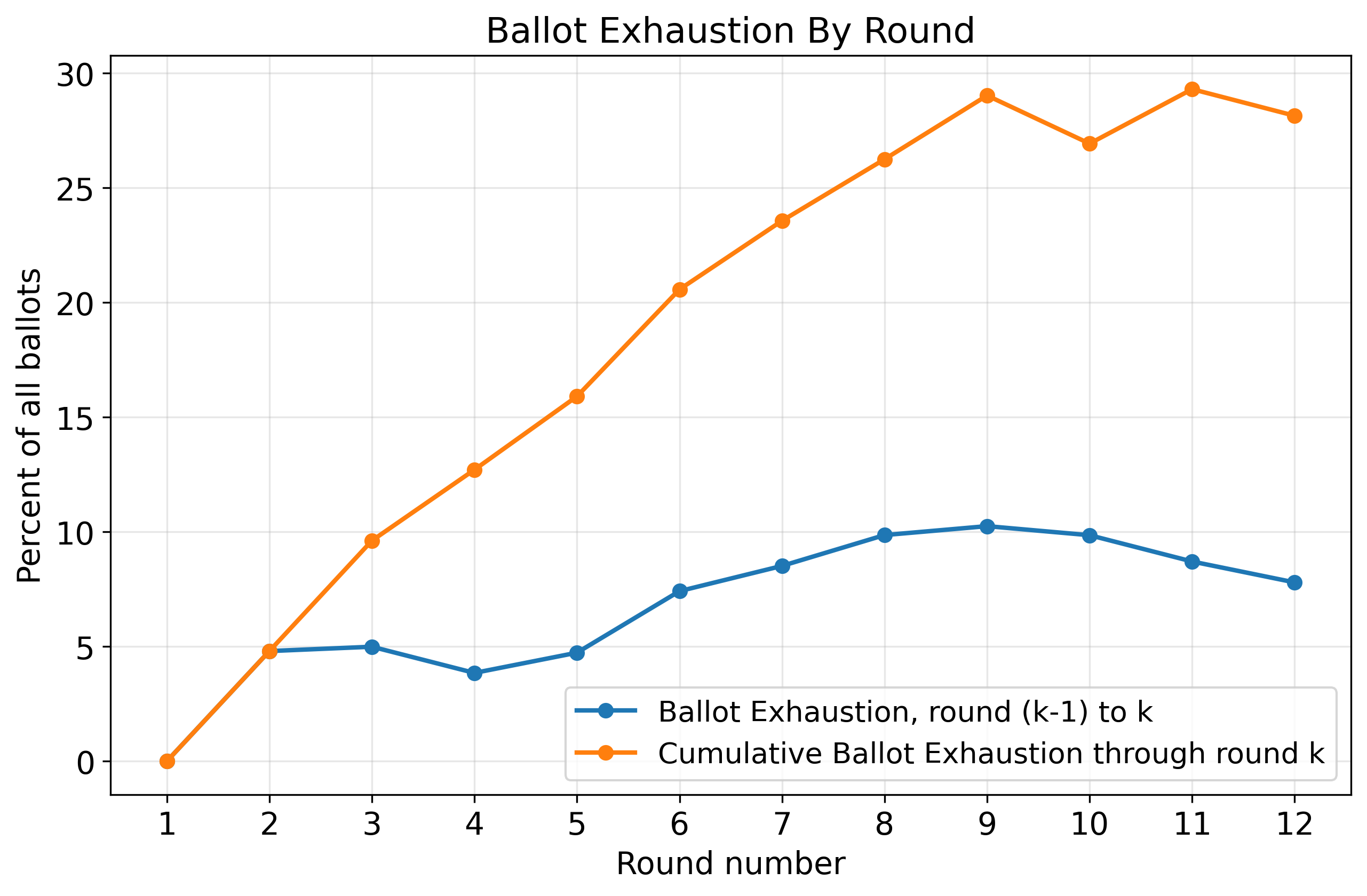}
\caption{The rate of ballot exhaustion by round across all elections which last at least $k$ rounds.}
\label{figure:exhaustion_by_round}
\end{figure}

\begin{figure}[h]
\includegraphics[width=130mm]{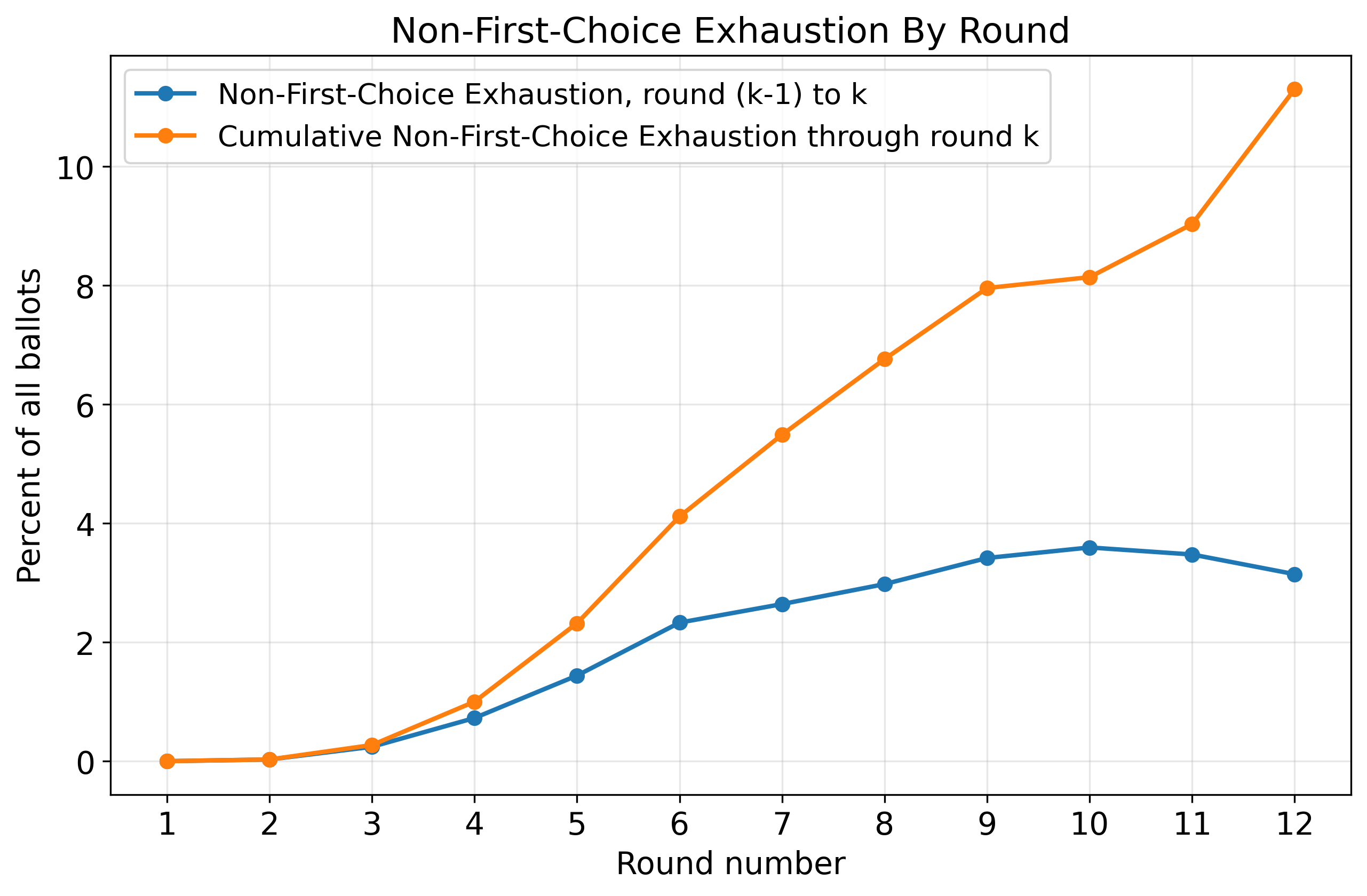}
\caption{The rate of non-first-choice ballot exhaustion by round across all elections which last at least $k$ rounds.}
\label{figure:exhaustion_by_round_non_first_choice}
\end{figure}

\begin{figure}[h]
\includegraphics[width=130mm]{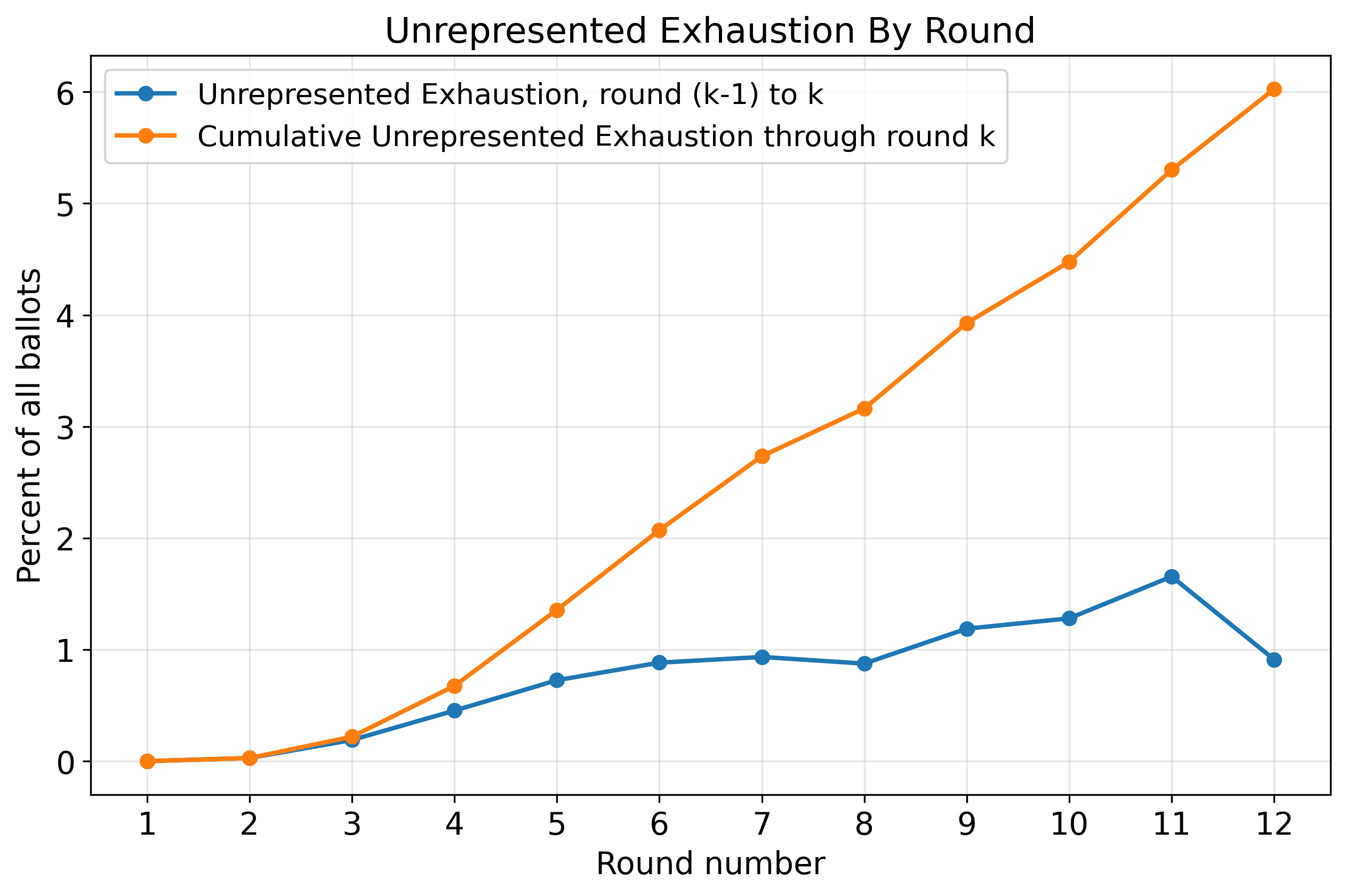}
\caption{The rate of unrepresented ballot exhaustion by round across all elections which last at least $k$ rounds.}
\label{figure:exhaustion_by_round_unrepresnted}
\end{figure}

\begin{figure}[h]
\includegraphics[width=130mm]{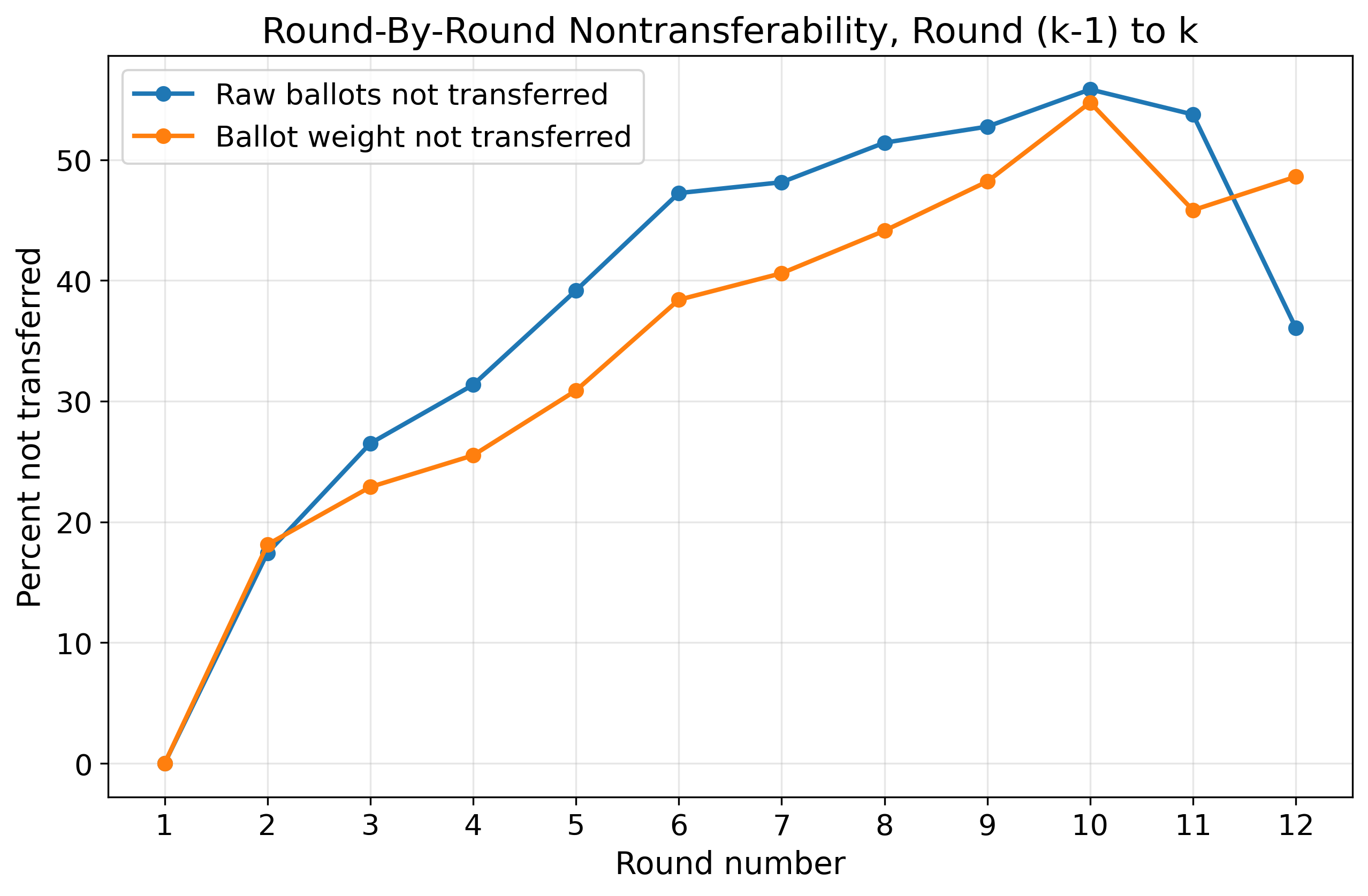}
\caption{The round-by-round percent of non-transferable ballots and non-transferable weight  across all elections which run for at least $k$ rounds.}
\label{figure:exhaustion_round_by_round}
\end{figure}

\begin{figure}[h]
 \includegraphics[width=100mm]{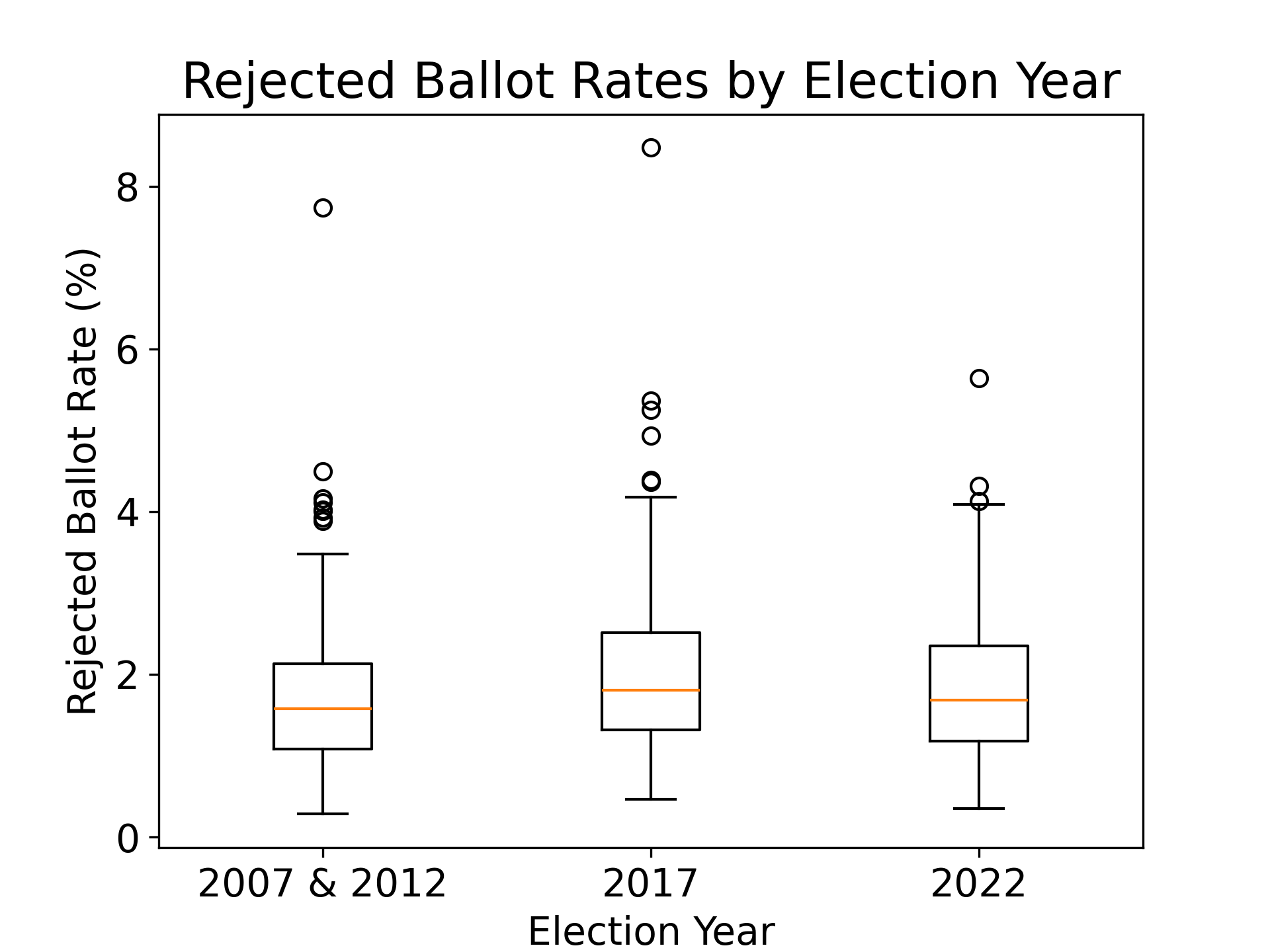}  
 \caption{Rejected ballot rates separated by election year. For 2007 we have elections only for the district of Glasgow City, and we group these elections with the 2012 elections. To make the figure more readable we omit one outlier election from 2017, a Dundee City council election with a rejected ballot rate of 12.3\%.}
 \label{figure:rejected_ballots}
\end{figure}

\end{document}